\documentclass[twocolumn]{autart}

\usepackage{algorithm}
\usepackage{algpseudocode}

\usepackage{amsmath}
\usepackage{amssymb}
\usepackage{accents} 
\usepackage{enumitem} 
\usepackage{natbib}
\usepackage{pgfplots}
\pgfplotsset{compat=newest}
\usetikzlibrary{arrows.meta, 
                calc, 
                external, 
                plotmarks, 
                shadows.blur, 
                shapes.geometric, 
                }
\usepackage{xcolor}
\usepackage{soul}

\begin{document}

\begin{frontmatter}
    \title{Data-Efficient Extremum-Seeking Control Using Kernel-Based Function Approximation\thanksref{footnoteinfo}}
    \thanks[footnoteinfo]{This publication is part of the project Digital Twin with project number P18-03 of the research program TTW Perspective which is (partly) financed by the Dutch Research Council (NWO). Corresponding author W. Weekers.}
    \author[1]{Wouter Weekers}\ead{w.weekers@tue.nl},
    \author[1]{Alessandro Saccon}\ead{a.saccon@tue.nl},
    \author[1]{Nathan van de Wouw}\ead{n.v.d.wouw@tue.nl}
    \address[1]{Eindhoven University of Technology, Department of Mechanical Engineering, 5600 MB, Eindhoven, The Netherlands}
    \begin{abstract}
        Existing extremum-seeking control (ESC) approaches typically rely on applying repeated perturbations to input parameters and performing measurements of the corresponding performance output. The required separation between the different timescales in the ESC loop means that performing these measurements can be time-consuming. Moreover, performing these measurements can be costly in practice, e.g., due to the use of resources. With these challenges in mind, it is desirable to reduce the number of measurements needed to optimize performance. Therefore, in this work, we present a sampled-data ESC approach aimed at reducing the number of measurements that need to be performed. In the proposed approach, we use input-output data obtained during regular operation of the extremum-seeking controller to construct online an approximation of the system's underlying cost function. By using this approximation to perform parameter updates when a decrease in the cost can be guaranteed, instead of performing additional measurements to perform this update, we make more efficient use of data collected during regular operation of the extremum-seeking controller. As a result, we indeed obtain a reduction in the required number of measurements to achieve optimization. We provide a stability analysis of the novel sampled-data ESC approach, and demonstrate the benefits of the synergy between kernel-based function approximation and standard ESC in simulation on a multi-input dynamical system.%
    \end{abstract}
\end{frontmatter}

\section{Introduction}
\label{sec:introduction}

Extremum-seeking control (ESC) is a data-driven control approach aimed at optimizing the steady-state performance of unknown dynamical systems. Existing ESC approaches \citep[see, e.g.,][]{Tan2010,Scheinker2024} are generally based on the assumption that there exists a time-invariant input-output map that uniquely relates constant system input parameters to the corresponding steady-state performance output. To optimize the steady-state performance, typically, a small perturbation signal (dither) is added to the input parameters, and the corresponding system output is measured. In this way, information about the system's underlying input-output map, such as its gradient, is obtained that is used in an optimization algorithm to steer the input parameters to their performance-optimizing values.

To ensure stable convergence of the input parameters to a small neighborhood of their performance-optimizing values, the different timescales in the ESC loop (the system dynamics, the dither, and optimizer dynamics) should be separated such that the measured performance output remains close to its steady-state solution and the effect of measurement errors on the dynamics of the ESC is small. In classical continuous-time ESC approaches \citep{Krstic2000,Tan2006,Tan2010,Scheinker2024}, for example, this timescale separation is achieved by choosing the frequency of the sinusoidal dither signal sufficiently low compared to the dominant timescale of the system dynamics, and the cut-off frequencies of the high- and low-pass filters used to estimate the gradient of the input-output map via demodulation sufficiently low compared to the dither frequency. However, due to the need for this timescale separation, measuring the (steady-state) performance output can be time-consuming. In addition, it typically causes the convergence rate of the input parameters to be low. Moreover, the required continuous or repeated application of dither and the required measurements of the corresponding performance output can be costly or undesirable in practical applications, e.g., due to the use of resources such as communication bandwidth or (raw) materials, or due to the transient system responses caused by the dither resulting in scrap or waste products.

Many continuous-time approaches have been proposed that improve the convergence rate in ESC. These approaches include Newton-based ESC \citep{Ghaffari2012} and fixed-time ESC \citep{Poveda2021_fixed_time} to speed up the learning dynamics, bounded ESC approaches that allow optimization even when the system is unstable \citep{Scheinker2016}, observer-based approaches \citep{Ryan2010,Gelbert2012,Guay2015,Haring2018} that improve the gradient estimate and allow eliminating the timescale associated with the high- and low-pass filters in classical approaches, or the use of curve fitting techniques to obtain local \citep{Hunnekens2014} and non-local \citep{Poveda2021} approximations of the input-output map for gradient estimation. However, most of these approaches still require timescale separation between the dither and the system dynamics, meaning that obtaining the measurements of the (steady-state) performance output can still be time-consuming. To address this time-consuming nature in a continuous-time ESC context, in \citet{VanKeulen2020} online approximations of the system dynamics are constructed that allow eliminating this separation of timescales between the system dynamics and the dither. Furthermore, in \citet{Oliveira2017}, a delay compensation approach is presented for the case where the time-consuming nature stems from a known delay. In a sampled-data context, the time-consuming nature is (partially) addressed in \citet{Poveda2017} by replacing the typically fixed waiting time between applying inputs and measuring the corresponding steady-state output (see, e.g., \citet{Teel2001,Kvaternik2011,Khong2013a,Khong2013b,Hazeleger2022}) by a variable waiting time.

In addition, most of the aforementioned approaches still require continuous or repeated application of dither and performing measurements of the corresponding system output, which as mentioned before, can be costly in practice. Exceptions are the approaches presented in \citet{Hunnekens2014} and \citet{Guay2015}. The former approach allows eliminating the dither signal altogether, while the latter allows eliminating the dither in some cases. However, the stability proof presented in \citet{Hunnekens2014} is limited to static systems, and in \citet{Guay2015} no clear conditions are given for when the dither can be omitted. In \citet{Rodrigues2022,Rodrigues2023}, the issue of the costly nature of performing measurements is partially addressed for applications where the costly nature stems from the use of communication bandwidth, by using ideas from event-triggered control. However, this approach does not address the costly nature of performing measurements in a general problem setting.

Therefore, the main contribution of this work is a novel ESC approach which addresses the challenge of both the time-consuming and the costly nature of performing measurements in practical applications, by reducing the number of input-output measurements that need to be performed. In particular, we adopt a sampled-data ESC setting as in, e.g., \citet{Teel2001, Hazeleger2022}, since this allows for possible extensions to global optimization algorithms as, e.g., in \citet{Khong2013a,Khong2013b} in the future. In this setting, the novel ESC approach uses the input-output data collected during operation of the extremum-seeking controller to construct online an approximation of the steady-state input-output map using kernel-based function approximation. When the approximation is sufficiently accurate in a region around the current optimizer state, it is used to perform a parameter update step without requiring additional measurements to be made. In this way, more efficient use is made of previously collected data, which reduces the number of measurements that need to be performed. Moreover, unlike the typically many measurements required for offline system identification aimed at model-based optimization, our online approach does not require additional measurement effort to construct the approximation since it uses only the data collected during regular operation of the extremum-seeking controller. A formal stability analysis that shows that our novel ESC approach guarantees (under appropriate tuning conditions) convergence to a neighborhood of the optimum is provided.

In our preliminary work \citep{Weekers2023}, we considered only the optimization of static maps. The additional contribution of the current work is therefore the extension of this approach to the more general case of dynamical systems, including a stability proof of the presented approach. Furthermore, we derive novel conditions for assessing whether the kernel-based approximation is sufficiently accurate to be used for a parameter update step. These conditions, which are based on solving convex optimization problems, are less conservative than the closed-form expressions used in \citet{Weekers2023}. Finally, we show the benefits of the proposed approach in simulation on a nonlinear dynamical system with multiple inputs.

The remainder of this paper is organized as follows. In Section~\ref{sec:problem_formulation}, we introduce the considered class of dynamical systems and formulate the optimization problem that we aim to solve. In Section~\ref{sec:data-efficient_ES}, we present the proposed approach. A stability analysis for the proposed ESC scheme is performed in Section~\ref{sec:stability_analysis}. In Section~\ref{sec:simulation_study}, the approach is demonstrated by means of a simulation example. Finally, conclusions are given in Section~\ref{sec:conclusion}. Throughout the paper, we use the following notation.
\begin{itemize}
    \item A continuous function $\rho : \mathbb{R}_{\geq 0} \to \mathbb{R}_{\geq 0}$ is of class $\mathcal{K}$ ($\rho \in \mathcal{K}$), if it is strictly increasing, and $\rho(0)=0$. If in addition $\rho(r) \to \infty$ as $r \to \infty$, then $\rho$ is of class $\mathcal{K}_\infty$ ($\rho \in \mathcal{K}_\infty$).
    \item A continuous function $\beta: \mathbb{R}_{\geq 0} \times \mathbb{R}_{\geq 0}\to \mathbb{R}_{\geq 0}$ is of class $\mathcal{K}\mathcal{L}$ $(\beta \in \mathcal{K}\mathcal{L})$ if, for each fixed $t$, $\beta(\cdot,t) \in \mathcal{K}$, and, for each fixed $s$, $\beta(s,\cdot)$ is decreasing and $\beta(s,t) \to 0$ as $t \to \infty$. 
    \item Let $\mathcal{X}$ be a Banach space with norm $\lVert \cdot \rVert$. Given any $\mathcal{Y} \subset \mathcal{X}$, and a point $x \in \mathcal{X}$, $\lVert x \rVert_{\mathcal{Y}} := \inf_{a \in \mathcal{Y}} \, \lVert x-a \rVert$ defines the distance of $x$ to the set $\mathcal{Y}$.
    \item Let $\mathcal{A} + \varepsilon \bar{\mathcal{B}} := \{x \in \mathcal{X} : \lVert x \rVert_{\mathcal{A}} \leq \varepsilon$\} be the set of all points within a distance $\varepsilon$ of $\mathcal{A}$, i.e., $\mathcal{A} + \varepsilon\bar{\mathcal{B}}$, with $\bar{\mathcal{B}}$ denoting the closed unit ball, is an $\varepsilon$-neighborhood of $\mathcal{A}$.
    \item The superscript $(\cdot)^+$ denotes update steps for discrete-time systems, e.g., $\theta_{k+1}\in F(\theta_k)$, with $k$ the discrete time index, is denoted as $ \theta^+\in F(\theta)$.
    \item The identity function is denoted by ${\rm id}(\cdot)$.
\end{itemize}
\section{Problem formulation}
\label{sec:problem_formulation}

In this section, we first introduce the class of dynamical systems that we consider. Next, we formulate the optimization problem that we aim to solve, and recall a classical sampled-data extremum-seeking control approach that can be used to solve this problem.

\subsection{Considered class of dynamical systems}
\label{sec:class_of_systems_and_optimization_problem}

We consider a class of nonlinear, possibly infinite-dimensional systems, according to the following definition.
\begin{defn}
    \label{def:Sigma_p}
    Let $\Sigma_p$ be a time-invariant system, whose state and input are denoted by $x \in \mathcal{X}$ and $\theta \in \mathbb{R}^{n_\theta}$, respectively. Here, $\mathcal{X}$ is a Banach space with norm $\lVert \cdot \rVert$. Given any initial state $x_0 \in \mathcal{X}$, and any constant input $\theta \in \mathbb{R}^{n_\theta}$, the state trajectory of $\Sigma_p$ starting at $x_0$ with constant input $\theta$ is denoted by $x(\cdot, x_0, \theta)$.
\end{defn}
We adopt the following assumption for the class of systems in Definition~\ref{def:Sigma_p}.
\begin{assum}
    \label{as:Sigma_p}
    Given a system $\Sigma_p$ as in Definition~\ref{def:Sigma_p} we assume that the following statements hold:
    \begin{enumerate}[label=(\roman*), ref=\theassum(\roman*)]
        \item For any constant input $\theta \in \mathbb{R}^{n_\theta}$, there exists a closed and nonempty set $\mathcal{A}(\theta) \in \mathcal{X}$ such that
        \begin{align}
            \underset{t \to \infty}{\lim} \, \lVert x(t, x_0, \theta) \rVert _{\mathcal{A}(\theta)} = 0,
        \end{align}
        i.e., for any constant input the system's trajectories converge to a global attractor which defines a, potentially set-valued, map $\mathcal{A}(\cdot)$ from $\mathbb{R}^{n_\theta}$ to subsets of $\mathcal{X}$. \label{as:global_attractor}
        \item There exists an unknown, continuous function $h:\mathcal{X} \to \mathbb{R}$ that maps  the state trajectory $x(\cdot, x_0, \theta)$ to the system output $y$, such that for any initial state $x_0 \in \mathcal{X}$ and any constant input $\theta \in \mathbb{R}^{n_\theta}$
        \begin{align}
            y(t) := h(x(t, x_0, \theta)) \label{eq:y_t}
        \end{align}
        with $h(x_1) = h(x_2)$ for any $x_1, x_2 \in \mathcal{A}(\theta)$. As a consequence, since $\mathcal{A}(\theta)$ is a global attractor, the unknown steady-state input-output map 
        \begin{align}
            f(\theta):= \underset{t \to \infty}{\lim} \: h(x(t, x_0, \theta)) \label{eq:definition_f}
        \end{align}
        is well-defined for any initial state $x_0 \in \mathcal{X}$ and any constant input $\theta \in \mathbb{R}^{n_\theta}$. \label{as:f_well-defined}
        \item The steady-state input-output map $f(\cdot)$ is continuously differentiable and takes its (global) minimum value in a non-empty, compact set $\mathcal{C}$, i.e., $f(\theta) > f(\theta^*)$  for all $\theta \in \mathbb{R}^{n_\theta} \backslash \mathcal{C}$ with $\theta^* \in \mathcal{C}$. Moreover, $\nabla f(\theta) = 0$ if and only if $\theta \in \mathcal{C}$. \label{as:minimum_f}
        \item For any $\Delta_\theta, \Delta_x \in \mathbb{R}_{>0}$, there exists an $L_h \in \mathbb{R}_{>0}$ such that
        \begin{align}
            \lVert h(x) - f(\theta) \rVert \leq L_h \lVert x \rVert_{\mathcal{A}(\theta)}
        \end{align}  
        holds for any input $\theta \in \mathbb{R}^{n_\theta}$ and state $x \in \mathcal{X}$ that satisfy $\lVert \theta \rVert_{\mathcal{C}} \leq \Delta_\theta$ and $\lVert x \rVert_{\mathcal{A}(\theta)} \leq \Delta_x$.  \label{as:Lipschitz_h}
        \item The map $\mathcal{A}(\cdot)$ is locally Lipschitz. That is, for any $\Delta_\theta \in \mathbb{R}_{> 0}$, there exists an $L_{\mathcal{A}} \in \mathbb{R}_{>0}$ such that
        \begin{align}
            \mathcal{A}(\theta_1) \subseteq \mathcal{A}(\theta_2) + L_{\mathcal{A}}\lVert \theta_1-\theta_2\rVert \bar{\mathcal{B}}
        \end{align}
        if $\max\{\lVert \theta_1 \rVert_{\mathcal{C}},\, \lVert \theta_2 \rVert_{\mathcal{C}}\}\leq \Delta_\theta$.\label{as:Lipschitz_A}
        \item Given any $\epsilon_1, \epsilon_2, \Delta_{x}, \Delta_\theta \in \mathbb{R}_{>0}$, there exists a time $T \in \mathbb{R}_{>0}$, called a waiting time, such that for any constant input $\theta \in \mathbb{R}^{n_\theta}$ and initial condition $x_0 \in \mathcal{X}$ that satisfy $\lVert \theta \rVert_{\mathcal{C}} \leq \Delta_\theta$ and $\lVert x_0 \rVert_{\mathcal{A}(\theta)} \leq \Delta_{x}$, it holds that
        \begin{align}
            \lVert x(t, x_0, \theta)\rVert_{\mathcal{A}(\theta)} \leq \epsilon_1\lVert x_0 \rVert_{\mathcal{A}(\theta)} + \epsilon_2
        \end{align}
        for all $t \geq T$. \label{as:sufficiently_long_T}
    \end{enumerate}
\end{assum}
For examples of common classes of systems that satisfy  Definition~\ref{def:Sigma_p} and Assumption~\ref{as:Sigma_p} see \citet[Section~2]{Khong2013b}.
\begin{rem}
    \label{rm:assumptions_plant}
    The class of systems satisfying Definition~\ref{def:Sigma_p} and Assumption~\ref{as:Sigma_p} considered here is largely in line with the classes of systems considered in, e.g., \citet{Teel2001}, \citet{Khong2013b} and \citet{Hazeleger2022}. The main differences are that: (i) in those works $f(\cdot)$ is only assumed to be locally Lipschitz instead of continuously differentiable with $\nabla f(\theta) = 0 \Leftrightarrow \theta \in \mathcal{C}$ (the reason for the stricter assumption made here will be made clear in Section~\ref{sec:kernel-based_extremum-seeking}); (ii) in \citet{Khong2013b} the inputs $\theta$ are restricted to a compact subset of $\mathbb{R}^{n_\theta}$ while here $\theta \in \mathbb{R}^{n_\theta}$; and (iii) in \citet{Hazeleger2022} the system has additional outputs related to measurable constraints, while no constraints are considered here. 
\end{rem}

\subsection{Problem formulation}
\label{sec:standard_extremum-seeking}

In the context of ESC, the input $\theta$ and output $y$ of the system $\Sigma_p$ can be seen as a vector of tunable system parameters and a to-be-optimized performance variable, respectively. The (constant) input and the output are related in steady state via the steady-state input-output map $f(\cdot)$ in \eqref{eq:definition_f}, which thus represents a steady-state cost function for the system $\Sigma_p$. However, the system dynamics and consequently the input-output map $f(\cdot)$ are unknown. The goal in ESC is therefore to solve the steady-state optimization problem
\begin{align}
    \theta^* := \operatorname*{\arg\min}_{\theta \in \mathbb{R}^{n_\theta}} f(\theta) \label{eq:theta_star}
\end{align}
solely on the basis of input-output data of the system $\Sigma_p$, i.e., only using knowledge on the applied input $\theta$ and the measured output $y$.

To solve the steady-state optimization problem \eqref{eq:theta_star}, we will focus in particular on sampled-data ESC approaches in which the optimization algorithm can be described by a difference inclusion (see also, e.g., \citet{Teel2001} and \citet{Hazeleger2022}). Extensions to a broader class of optimization algorithms, such as in \citet{Khong2013a,Khong2013b} for example, are left as future work. The class of optimization algorithms that we consider can thus be described by
\begin{align}
    \Sigma : \quad  \theta^+ \in F(\theta, Y(\theta)), \label{eq:Sigma}
\end{align}
where $F : \mathbb{R}^{n_\theta} \times \mathbb{R}^{n_v} \to \mathbb{R}^{n_\theta}$ is a set-valued map whose input-output behavior depends on tunable parameters of the optimization algorithm, and the updated parameters $\theta^+$ can be any element of the set. Furthermore, the map $Y : \mathbb{R}^{n_\theta} \to \mathbb{R}^{n_v}$ defined as 
\begin{align}
    Y(\theta) := \begin{bmatrix}
        f(\theta + v_1(\theta))&
        \cdots&
        f(\theta + v_{n_v}(\theta))
    \end{bmatrix}^\intercal \label{eq:Y}
\end{align}
with dither functions $v_j(\cdot)$, $j = 1, 2, \dots, n_v$, and input-output map $f(\cdot)$, maps inputs $\theta$ to vectors containing information regarding (the gradient of) $f(\cdot)$ near $\theta$.

However, as mentioned before, the input-output map $f(\cdot)$ is unknown, and thus the elements of $Y(\theta)$ in \eqref{eq:Y} can only be evaluated via measurements of the system output $y$. We recall how these measurements are obtained during what we will refer to as a \textit{standard} parameter update step: a single iteration of the ESC algorithm considered in, e.g., \citet{Hazeleger2022} (taking the number of constraints equal to zero) and \citet{Teel2001}. Such an iteration requires $n_v$ sequential measurements in which a constant input including dither $v_j(\cdot)$ is applied to the system for a time duration $T$, and the output is sampled at the end of the time interval. That is,
\begin{align}
    \theta(t) = \widetilde{\theta}_i & := \widehat{\theta}_k + v_j(\widehat{\theta}_k)\nonumber\\
    & \forall t\in [t_{0,k} + (j - 1)T,\, t_{0,k} + jT)\label{eq:theta_tilde}
\end{align}
and
\begin{align}
    \widetilde{y}_i := y(t_{0,k} + jT) \label{eq:sampler_y}
\end{align}
for each perturbation in the interval with index $j=1,2,\dots,n_v$ during the current optimizer iteration $k$. Here, $i = N_k + j$ with $N_k$ the number of measurements performed up to iteration $k$, and $t_{0,k}$ and $\widehat{\theta}_k$ denote, respectively, the time and optimizer state at the start of iteration $k$. Based on the collected samples $\widetilde{y}_i$, the optimizer state $\widehat{\theta}_k$ is updated as in \eqref{eq:Sigma}-\eqref{eq:Y}, i.e.,
\begin{align}
    \widehat{\theta}_{k+1}\in F(\widehat{\theta}_k,\widetilde{Y}(\widehat{\theta}_k)), \label{eq:Sigma_perturbed}
\end{align}
where $\widehat{\theta}_{k+1}$ can be any element of the set induced by $F$ and
        \begin{align}
            \widetilde{Y}(\widehat{\theta}_k):=
            \begin{bmatrix}
                \widetilde{y}_{N_k + 1} & \cdots & \widetilde{y}_{N_k + n_v}
            \end{bmatrix}^\intercal\label{eq:Ytilde}
        \end{align}
is an approximation of $Y(\widehat{\theta}_k)$ with $Y(\cdot)$ as in \eqref{eq:Y}, since the system is not fully in steady state at the moment the samples $\widetilde{y}_{N_k+1},\widetilde{y}_{N_k+2},\dots,\widetilde{y}_{N_k+n_v}$ are taken. Note that $\widetilde{Y}$ depends on $\widehat{\theta}_k$ through the samples of $y$ (cf. \eqref{eq:y_t}).

To analyze convergence properties of the optimization algorithm to the set of minimizers $\mathcal{C}$, and the effect that the discrepancies between $Y(\cdot)$ in \eqref{eq:Y} and $\widetilde{Y}(\cdot)$ in \eqref{eq:Ytilde} have on these convergence properties, we adopt the following assumptions on the optimization algorithm \eqref{eq:Sigma}-\eqref{eq:Y} and the dither functions $v_j(\cdot)$.
\begin{assum}
    \label{as:Sigma}
    Given the optimization algorithm and dither functions $v_j(\cdot)$ as in \eqref{eq:Sigma}-\eqref{eq:Y}, we assume that the following statements hold:
    \begin{enumerate}[label=(\roman*), ref=\theassum(\roman*)]
        \item For each input $\theta \in \mathbb{R}^{n_\theta}$, the set $F(\theta, Y(\theta))$ in \eqref{eq:Sigma} is nonempty and compact.  Moreover, $F$ is an upper semi-continuous function of $\theta$. \label{as:nonempty_compact_F}
        \item  There exist class-$\mathcal{K}_\infty$ functions $\omega_1$, $\omega_2$, and $\rho$, and a nonnegative constant $\delta \in \mathbb{R}_{\geq 0}$, such that, for any $\theta \in \mathbb{R}^{n_\theta}$ and $\theta^* \in \mathcal{C}$,
        \begin{equation}
            \omega_1(\lVert \theta \rVert_{\mathcal{C}}) \leq V(\theta) \leq \omega_2(\lVert \theta \rVert_{\mathcal{C}} ) \label{eq:omega1_f_omega_2}
        \end{equation}
        and \label{as:DV_standard}
        \begin{align}
            \underset{\theta^+ \in F(\theta, Y(\theta))}{\max} \: V(\theta^+) - V(\theta) \leq -\rho(V(\theta)) + \delta \label{eq:Delta_f}
        \end{align}
        with
        \begin{align}
            V(\theta) := f(\theta) - f(\theta^*). \label{eq:definition_V}
        \end{align}
        \item The constant $\delta \in \mathbb{R}_{\geq 0}$ in \eqref{eq:Delta_f} can be made arbitrarily small by tuning parameters of the optimization algorithm $F$ in \eqref{eq:Sigma}.\label{as:small_delta_standard}
        \item For any $\Delta_\theta, \Delta_Y \in \mathbb{R}_{> 0}$, there exists an $L_Y \in \mathbb{R}_{> 0}$ such that if $\lVert \theta \rVert_{\mathcal{C}} \leq \Delta_\theta$ and $\lVert \widetilde{Y}(\theta) \rVert \leq \Delta_Y$, then 
        \begin{align}
            \lVert \theta_{\smash{\widetilde{F}}} - \theta_F \rVert \leq L_Y \lVert \widetilde{Y}(\theta)-Y(\theta) \rVert,
        \end{align} 
        where $\theta_{\smash{\widetilde{F}}}$ denotes any element of the set generated by $F(\theta, \widetilde{Y}(\theta))$, and $\theta_F$ denotes its closest point in the set $F(\theta, Y(\theta))$. \label{as:Lipschitz_F}
        \item For any $\Delta_\theta \in \mathbb{R}_{> 0}$, there exist constants $M_v, c_v \in \mathbb{R}_{\geq 0}$ such that  
        \begin{align}
            \lVert v_j(\theta) \rVert \leq M_v \lVert \theta \rVert_{\mathcal{C}} + c_v, \quad j = 1, 2, \dots, n_v,
        \end{align}
        for any input $\theta$ such that $\lVert \theta \rVert_{\mathcal{C}} \leq \Delta_\theta$. \label{as:dither_bound}
    \end{enumerate}
\end{assum}
\begin{rem}
    \label{rm:assumptions_optimizer}
    Assumption~\ref{as:Sigma} is largely aligned with the assumptions on the optimizer made in \citet{Teel2001} and \citet{Hazeleger2022} (see also the remark in \citet[Section~6.2]{Khong2013b} about the assumptions in \citet{Teel2001}). The main differences are that: (i) in those works $V(\theta)$ in \eqref{eq:omega1_f_omega_2} and \eqref{eq:Delta_f} is a general locally Lipschitz function $V(\theta)$ instead of being explicitly defined as in \eqref{eq:definition_V} (the reason for our stricter assumption will be made clear in Section~\ref{sec:stability_analysis}); and (ii) in \citet{Hazeleger2022} additional terms are present in the assumptions that relate to measurable constraints which are not considered here.
\end{rem}
 Note that each standard update step \eqref{eq:theta_tilde}-\eqref{eq:Ytilde} requires $n_v$ sequential experiments in which the input-output map $f(\cdot)$ is approximately evaluated for the inputs $\widetilde{\theta}_i$ (optimizer state plus dither) via the measured outputs $\widetilde{y}_i$. Furthermore, the data collected during these experiments are typically only used for a single algorithm update step, after which they are discarded. Given the potentially time-consuming and costly nature of performing these sequential measurements, we present in the next section an approach in which we use the input-output data $\mathcal{D}_k = \{(\widetilde{\theta}_i, \widetilde{y}_i) \mid i = 1, 2, \dots, N_k\}$ collected during such standard update steps (i.e., during regular operation of the extremum-seeking controller) to construct online an approximation of the input-output map $f(\cdot)$. When this approximation is sufficiently accurate in a sense that we will clarify later, we use it to determine a search direction and suitable optimizer gain for a gradient-based optimization step without the need for additional measurements.
\section{Data-efficient extremum-seeking using function approximation}
\label{sec:data-efficient_ES}

This section consists of three parts. In Section~\ref{sec:kernel-based_extremum-seeking}, we outline how an approximation of the steady-state input-output map can be used to perform a parameter update step (instead of a standard update step as in \eqref{eq:theta_tilde}-\eqref{eq:Ytilde}) whenever a decrease in cost can be guaranteed, to reduce the number of measurements needed to optimize performance. In Section~\ref{sec:kernel-based_approximation}, we discuss how such an approximation can be obtained from data collected during regular operation of the extremum-seeking controller. Finally, in Section~\ref{sec:approximation_accuracy} we discuss how the bounds used to assess whether a decrease can be guaranteed are obtained and present the full approach.

\subsection{Parameter updates based on cost function approximation}
\label{sec:kernel-based_extremum-seeking}

Suppose that a continuously differentiable approximation $m(\cdot)$ of the steady-state input-output map $f(\cdot)$ described in Section~\ref{sec:problem_formulation} is available. We will describe in Section~\ref{sec:kernel-based_approximation} how such an approximation can be obtained on the basis of input-output data collected during regular operation of the extremum-seeking controller. Since $m(\cdot)$ approximates $f(\cdot)$, its gradient $\nabla m(\theta)$ also forms an approximation of the gradient of $f(\theta)$. Hence, we can use it for a gradient-based update of $\theta$ aimed at minimizing $f(\theta)$. That is, we can use $\nabla m(\theta)$ for an update of the form
\begin{align}
    \theta^+ = \theta - \mu\nabla m(\theta) \label{eq:theta_plus_kb}
\end{align}
with $\mu \in \mathbb{R}_{>0}$ a suitably chosen optimizer gain. Unlike the standard parameter update \eqref{eq:theta_tilde}-\eqref{eq:Ytilde}, an update of the form \eqref{eq:theta_plus_kb} does not require additional measurements to be performed. Hence, replacing standard updates by updates of the form \eqref{eq:theta_plus_kb} reduces the total number of (time-consuming and costly) measurements required to solve the optimization problem in \eqref{eq:theta_star}.

A common approach for choosing the optimizer gain $\mu$ in \eqref{eq:theta_plus_kb}, used in static optimization, is to perform a backtracking line search until the so-called Armijo condition is satisfied (see, e.g., \citet[Chapter~3]{Nocedal2006}). This condition is given by
\begin{align}
    f(\theta^+) \leq f(\theta) - c \mu \nabla f(\theta)^\intercal \nabla m(\theta),\label{eq:Armijo}
\end{align}
where $\theta^+$ is as in \eqref{eq:theta_plus_kb} and $c \in (0,1)$ is a standard parameter that is typically chosen in the order of $10^{-4}$ (see, e.g., \citet[Chapter~3]{Nocedal2006}). If $-\nabla m(\theta)$ is a descent direction of $f(\theta)$, i.e., if
\begin{equation}
   \nabla f(\theta)^\intercal \nabla m(\theta) > 0, \label{eq:nabla_m_descent}
\end{equation}
then satisfaction of the Armijo condition \eqref{eq:Armijo} guarantees a decrease in $f(\theta)$ as a result of the update step. Furthermore, if \eqref{eq:Armijo} and \eqref{eq:nabla_m_descent} are satisfied for all update steps, $\theta$ converges to the optimizer (provided that Assumption~\ref{as:minimum_f} holds, cf. Remark~\ref{rm:assumptions_plant}).

However, in contrast to static optimization, where the cost function (and its gradient) are typically known, the steady-state input-output map $f(\cdot)$ is unknown in the context of ESC, and can only be evaluated approximately by performing measurements. Moreover, its gradient $\nabla f(\cdot)$ is unavailable, and can only be approximated by performing multiple (approximate) evaluations of $f(\cdot)$. Hence, verifying satisfaction of \eqref{eq:Armijo} and \eqref{eq:nabla_m_descent} (to determine if $-\nabla m(\theta)$ is a descent direction and to find a suitable $\mu$) would require performing multiple measurements, potentially adversely affecting the benefit of performing update steps of the form \eqref{eq:theta_plus_kb} in the first place. Yet, we will show in Section~\ref{sec:approximation_accuracy} that when the approximation $m(\cdot)$ is constructed in the way that we will describe in Section~\ref{sec:kernel-based_approximation}, it is possible to derive bounds $\underaccent{\bar}{b}(\theta)$ and $\bar{b}(\theta,\mu)$ that can be evaluated without performing measurements and that satisfy
\begin{equation}
    \nabla f(\theta)^\intercal \nabla m(\theta) \geq \underaccent{\bar}{b}(\theta) \label{eq:nabla_m_descent_bound}
\end{equation}
and
\begin{equation}
    f(\theta^+) - f(\theta) + c \mu \nabla f(\theta)^\intercal \nabla m(\theta) \leq \bar{b}(\theta,\mu)\label{eq:Armijo_bound}
\end{equation} 
with $\theta^+$ as in \eqref{eq:theta_plus_kb}. Hence, we can use these bounds to verify whether \eqref{eq:Armijo} and \eqref{eq:nabla_m_descent} are satisfied, without performing measurements, by checking whether $\underaccent{\bar}{b}(\theta) > 0$ and $\bar{b}(\theta,\mu) \leq 0$.

\begin{figure*}[tb!]
    \centering
    \includegraphics[width=\linewidth]{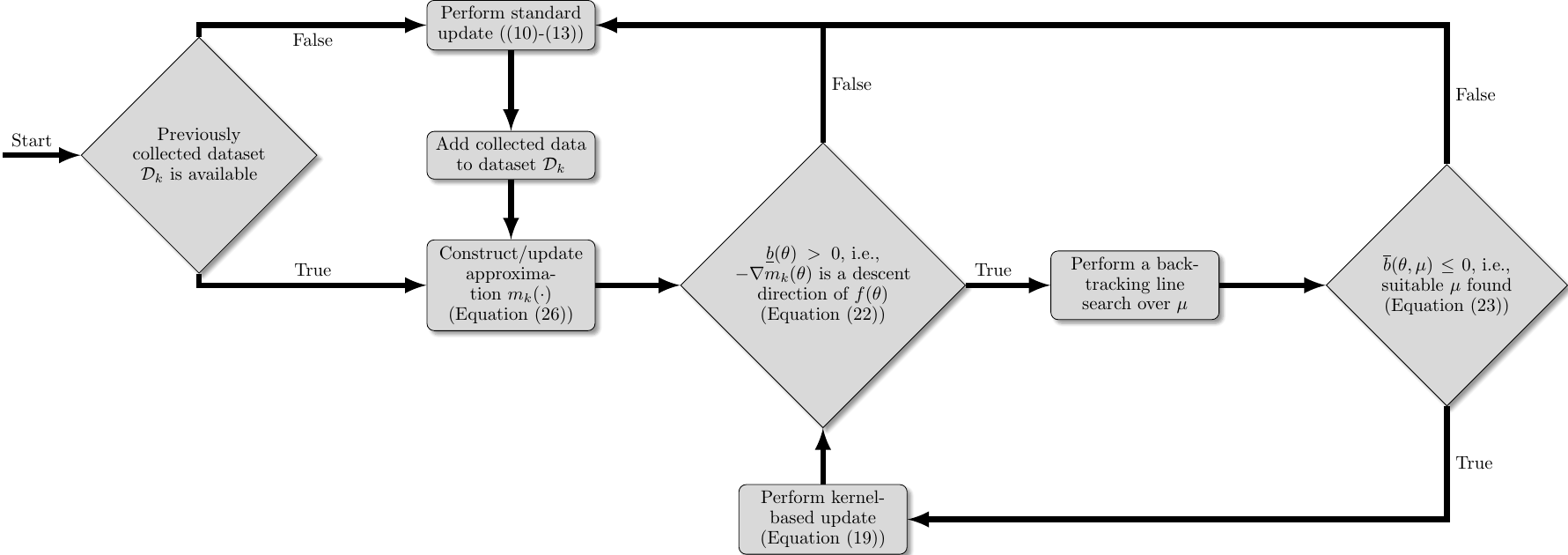}
    \caption{Block scheme illustrating kernel-based extremum-seeking control.}
    \label{fig:block_scheme}
\end{figure*}

Given the discussion above, we provide in Figure~\ref{fig:block_scheme} an illustration of our approach aimed at reducing the total number of measurements needed to solve the optimization problem in \eqref{eq:theta_star} in the context of ESC, which we briefly describe next. More details will be provided in Section~\ref{sec:approximation_accuracy}.

Suppose that a dataset $\mathcal{D}_k = \{(\widetilde{\theta}_i, \widetilde{y}_i) \mid i = 1,2,\dots, N_k \}$ of $N_k$ inputs $\widetilde{\theta}_i$ and corresponding measured outputs $\widetilde{y}_i$, e.g., as in \eqref{eq:theta_tilde}  and \eqref{eq:sampler_y}, respectively, is available. If such a dataset is not available a priori, we simply obtain it during a single standard parameter update step as in \eqref{eq:theta_tilde}-\eqref{eq:Ytilde}. From $\mathcal{D}_k$, we construct an approximation $m_k(\cdot)$ of the steady-state input-output map $f(\cdot)$ using kernel-based function approximation, as we will explain in Section~\ref{sec:kernel-based_approximation}. Then, for the current input $\theta$, we verify whether $-\nabla m_k(\theta)$ is a descent direction of $f(\theta)$ by checking whether $\underaccent{\bar}{b}(\theta) > 0$ (cf. \eqref{eq:nabla_m_descent} and \eqref{eq:nabla_m_descent_bound}). If this is the case, we perform a backtracking line search over $\mu$ to obtain a suitable optimizer gain $\mu$ for which $\bar{b}(\theta, \mu) \leq 0$, such that the Armijo condition \eqref{eq:Armijo} is guaranteed to be satisfied (cf. \eqref{eq:Armijo_bound}). If such a $\mu$ is found, we perform the parameter update step as in \eqref{eq:theta_plus_kb}, which does not require additional measurements. In case $\underaccent{\bar}{b}(\theta) \leq 0$ or no $\mu$ satisfying $\bar{b}(\theta,\mu) \leq 0$ has been found during the line search, we instead perform a standard parameter update step as in \eqref{eq:theta_tilde}-\eqref{eq:Ytilde} to update the parameters $\theta$. Subsequently, we add the input-output data collected during this standard update step to the dataset $\mathcal{D}_{k+1}$ to construct an updated approximation $m_{k+1}(\cdot)$ during the next iteration of the ESC algorithm.
 
We will refer to the approach outlined above as \emph{kernel-based extremum-seeking control} (KB-ESC). Here, `kernel-based' refers to the fact that a kernel-based approximation $m_k(\cdot)$ of the input-output map $f(\cdot)$ is constructed and used for some parameter update steps. Furthermore, we will refer to parameter update steps as in \eqref{eq:theta_plus_kb} as \emph{kernel-based} update steps, to distinguish them from the standard parameter update steps in \eqref{eq:theta_tilde}-\eqref{eq:Ytilde}.

\subsection{Approximating the steady-state input-output map using kernels}
\label{sec:kernel-based_approximation}

As mentioned in Section~\ref{sec:kernel-based_extremum-seeking}, we choose to construct the approximation $m_k(\cdot)$ of the input-output map $f(\cdot)$ using kernel-based function approximation. We take a kernel to be a continuous, symmetric function $\kappa : \mathbb{R}^{n_\theta} \times \mathbb{R}^{n_\theta} \to \mathbb{R}$ that is positive definite according to the following definition (see, e.g., \citet[Definition~3]{Scholkopf2001}). 
\begin{defn}
    \label{def:positive_definite_kernel}
    A function $\kappa : \mathbb{R}^{n_\theta} \times \mathbb{R}^{n_\theta} \to \mathbb{R}$ is called positive definite if for any $N \in \mathbb{N}$ and any set of inputs $\Omega = \{\theta_1, \dots, \theta_N\}$, the $N \times N$ matrix $K_{\Omega\Omega}$, whose $(i,j)$-th element is given by $\kappa(\theta_i, \theta_j)$, is positive semi-definite.
\end{defn}
By the Moore-Aronszajn theorem \citep{Aronszajn1950}, any kernel satisfying Definition~\ref{def:positive_definite_kernel} has a unique reproducing kernel Hilbert space (RKHS) of functions associated with it. This RKHS, which we will denote by $\mathcal{H}$, is the completion of the space of functions of the form $g(\cdot) = \sum_{i=1}^N \alpha_i \kappa(\cdot, \theta_i)$ with respect to the norm $\lVert g \rVert := \sqrt{\langle g, g \rangle}$ corresponding to the inner product $\langle g_1, g_2 \rangle =  \left\langle \sum_{i=1}^N \alpha_i\kappa(\cdot,\theta_i), \sum_{j=1}^M\beta_j\kappa(\cdot,\theta_j^\prime) \right\rangle = \sum_{i=1}^N \sum_{j=1}^M \alpha_i \beta_j \kappa(\theta_i,\theta_j^\prime)$. Here, $N,M\in\mathbb{N}$, $\alpha_1, \dots,\alpha_N, \allowbreak \beta_1, \dots, \beta_M\in\mathbb{R}$, and $\theta_1, \dots, \theta_N, \theta^\prime_1, \dots, \theta^\prime_M \in \mathbb{R}^{n_\theta}$. Note that $\lVert g \rVert^2$ takes the quadratic form
\begin{align}
    \lVert g \rVert^2 &= \langle g, g \rangle = \left\langle \sum_{i=1}^N \alpha_i \kappa(\cdot, \theta_i), \sum_{j=1}^N \alpha_j \kappa(\cdot, \theta_j) \right\rangle \nonumber\\
    &= \sum_{i=1}^N \sum_{j=1}^N \alpha_i \alpha_j \kappa(\theta_i, \theta_j) = \alpha^\intercal K_{\Omega\Omega} \alpha\label{eq:g_norm}
\end{align}
with $K_{\Omega\Omega}$ as in Definition~\ref{def:positive_definite_kernel} and $\alpha = [\alpha_1 \, \cdots \, \alpha_N]^\intercal$ a vector of weights.

We adopt the following two assumptions (see also \citet[Assumptions~1 and 2]{Scharnhorst2023}).
\begin{assum}
\label{as:bounded_measurment_error}
An upper bound $\bar{\delta}\in\mathbb{R}_{\geq 0}$ is known, such that $\lvert f(\widetilde{\theta}_i) - \widetilde{y}_i \rvert \leq \bar{\delta}$ for all measured outputs $\widetilde{y}_i$, $i=1,2,\dots,N_k$. That is, for every measured output $\widetilde{y}_i$, its distance to the corresponding steady-state value $f(\widetilde{\theta}_i)$ is at most $\bar{\delta}$.
\end{assum}
\begin{assum}
\label{as:true_function_in_RKHS}
Given a kernel $\kappa$, the steady-state input-output map $f(\cdot)$ belongs to its corresponding RKHS $\mathcal{H}$, and an upper bound $\Gamma$ on its norm $\lVert f \rVert$ is known.
\end{assum}
These assumptions will allow us to derive the bounds $\underaccent{\bar}{b}(\theta)$ and $\bar{b}(\theta,\mu)$ mentioned in Section~\ref{sec:kernel-based_extremum-seeking}, which we will use to verify, without performing measurements, whether $-\nabla m_k(\theta)$ is a descent direction of $f(\theta)$ and whether the Armijo condition \eqref{eq:Armijo} is satisfied for a given $\mu$. That is, whether we can perform a kernel-based update step with this $\mu$.
\begin{rem}
    Note that, instead of requiring $\bar{\delta}$ to be known in Assumption~\ref{as:bounded_measurment_error}, it can also be considered a design parameter since, given Assumptions~\ref{as:Lipschitz_h} and \ref{as:sufficiently_long_T}, Assumption~\ref{as:bounded_measurment_error} can be satisfied for any $\bar{\delta}\in\mathbb{R}_{>0}$ by choosing a sufficiently long waiting time $T$. Furthermore, the choice of kernel $\kappa$ in Assumption~\ref{as:true_function_in_RKHS} allows including prior knowledge that might be available about the input-output map $f(\cdot)$. In case no prior knowledge about this input-output map is available, universal kernels such as the squared exponential kernel 
    \begin{align}
        \kappa(\theta, \theta^\prime) = \exp{(-\lVert \theta - \theta^\prime \rVert^2/(2\sigma^2))} \label{eq:SE_kernel}
    \end{align}
    with $\sigma$ a tunable length scale, could be chosen since such kernels have the ability to approximate any continuous function arbitrarily closely on a compact set (see, e.g., \cite{Micchelli2006}). The bound $\Gamma$ in Assumption~\ref{as:true_function_in_RKHS} can be seen as a design parameter, which should be chosen large enough to guarantee that $\Gamma \geq \lVert f \rVert$, similar to how in standard sampled-data ESC the waiting time $T$ is a design parameter that should be chosen large enough to guarantee the system output $y$ is close to being in steady state when the output is measured.
\end{rem}

Given Assumptions~\ref{as:bounded_measurment_error} and \ref{as:true_function_in_RKHS}, we construct the approximation $m_k(\cdot)$ on the basis of data $\mathcal{D}_k = \{(\widetilde{\theta}_i, \widetilde{y}_i) \mid i = 1,2, \dots, N_k \}$ collected, e.g., during standard update steps \eqref{eq:theta_tilde}-\eqref{eq:Ytilde}, by solving the optimization problem
\begin{subequations}
    \label{eq:m_k_eSVR}
    \begin{align}
        m_k(\cdot) = \operatorname*{\arg\min}_{m \in \mathcal{H}} &\: \lVert m \rVert^2\\
        \text{s.t.} &\: \lvert m(\widetilde{\theta}_i) - \widetilde{y}_i \rvert \leq \bar{\delta} \nonumber\\
        &\quad \forall i\in\{1,2,\dots,N_k\},
    \end{align}
\end{subequations}
i.e., $m_k(\cdot)$ is the function in $\mathcal{H}$ with the smallest norm, that differs at most $\bar{\delta}$ from the data in $\mathcal{D}_k$. 

Note that, by Assumptions~\ref{as:bounded_measurment_error} and \ref{as:true_function_in_RKHS}, the steady-state input-output map $f(\cdot)$ is a feasible solution to \eqref{eq:m_k_eSVR}, and thus \eqref{eq:m_k_eSVR} is guaranteed to have a solution. 

By the representer theorem (see, e.g., \citet[Theorem~1]{Scholkopf2001}) the solution to \eqref{eq:m_k_eSVR} has the form 
\begin{equation}
    m_k(\cdot) = \sum_{i=1}^{N_k} \alpha_i \kappa(\cdot, \widetilde{\theta}_i) = K_{\Omega}(\cdot) \alpha,\label{eq:m_k_sum}
\end{equation}
where $K_\Omega(\cdot) := [\kappa(\cdot, \widetilde{\theta}_1) \, \kappa(\cdot, \widetilde{\theta}_2) \,  \cdots \, \kappa(\cdot, \widetilde{\theta}_{N_k})]$ is a row vector of kernel functions centered at the inputs $\Omega = \{\widetilde{\theta}_1, \widetilde{\theta}_2, \dots, \widetilde{\theta}_{N_k}\}$. Hence, substituting a solution $m(\cdot) = K_{\Omega}(\cdot) \alpha$ in \eqref{eq:m_k_eSVR}, and using \eqref{eq:g_norm}, shows that every time a new dataset $\mathcal{D}_k$ is obtained by adding new data pairs $(\widetilde{\theta}, \widetilde{y})$ to the previous dataset, a new approximation $m_k(\cdot)$ can be constructed by obtaining a new weight vector $\alpha\in\mathbb{R}^{N_k}$ by simply solving the quadratic programming problem
\begin{subequations}
    \label{eq:QP_alpha}
    \begin{align}
        \alpha = \operatorname*{\arg\min}_{\bar{\alpha} \in \mathbb{R}^{N_k}} \: & \bar{\alpha}^\intercal K_{\Omega\Omega} \bar{\alpha}\\
        \text{s.t.} \: & \lvert K_\Omega(\widetilde{\theta}_i) \bar{\alpha} - \widetilde{y}_i \rvert\leq \bar{\delta} \nonumber\\
        & \quad \forall i \in \{1,\dots,N_k\}
    \end{align}
\end{subequations}
and using \eqref{eq:m_k_sum}. Furthermore, since the weight vector $\alpha$ only depends on the dataset $\mathcal{D}_k$ and the chosen kernel $\kappa$, the gradient of the approximation $m_k(\cdot)$ is readily available as (cf. \eqref{eq:m_k_sum})
\begin{align}
    \nabla m_k(\cdot) &= \sum_{i=1}^{N_k} \alpha_i \begin{bmatrix} D^{(e_1, 0)}\kappa(\cdot, \widetilde{\theta}_i) & \cdots & D^{(e_{n_\theta}, 0)}\kappa(\cdot,\widetilde{\theta}_i) \end{bmatrix}^\intercal \nonumber\\
    &= \nabla K_\Omega(\cdot) \alpha\label{eq:nabla_m}
\end{align}
with $e_i \in \mathbb{R}^{n_\theta}$ a vector with $i$-th element equal to one and other elements equal to zero. In \eqref{eq:nabla_m}, we used the notation
\begin{align}
    D&^{(a,b)}\kappa(\theta, \theta^\prime) \nonumber\\
    & := \frac{\partial^{a_1 + \cdots + a_{n_\theta} + b_1 + \cdots + b_{n_\theta}}}{\partial \theta^{a_1}_1 \cdots \partial \theta^{a_{n_\theta}}_{n_\theta} \partial (\theta^\prime_1)^{b_1} \cdots \partial (\theta^\prime_{n_\theta})^{b_{n_\theta}}}\kappa(\theta, \theta^\prime)\label{eq:kernel_derivatives}
\end{align}
for partial derivatives of the kernel $\kappa$ (assuming that $\kappa$ is twice continuously differentiable), and
\begin{align}
    \nabla K_\Omega(\cdot) := \begin{bmatrix}
        D^{(e_1, 0)}\kappa(\cdot, \widetilde{\theta}_1) & \cdots & D^{(e_1, 0)}\kappa(\cdot, \widetilde{\theta}_{N_k})\\
        \vdots & \ddots & \vdots\\
        D^{(e_{n_\theta}, 0)}\kappa(\cdot, \widetilde{\theta}_1) & \cdots & D^{(e_{n_\theta}, 0)}\kappa(\cdot, \widetilde{\theta}_{N_k})\end{bmatrix} \label{eq:nabla_K_Omega}
\end{align}
denotes a matrix of partial derivatives of the kernel functions centered at the inputs $\Omega = \{\widetilde{\theta}_1, \widetilde{\theta}_2, \dots, \widetilde{\theta}_{N_k}\}$.

\subsection{Guaranteeing a decrease in cost during kernel-based update steps}
\label{sec:approximation_accuracy}
Given an approximation $m_k(\cdot)$ of the input-output map $f(\cdot)$ on the basis of a dataset $\mathcal{D}_k$ as in \eqref{eq:m_k_eSVR}, it remains to be assessed whether a kernel-based parameter update step \eqref{eq:theta_plus_kb} can be performed. To this end, as explained in Section~\ref{sec:kernel-based_extremum-seeking}, we need to determine whether $-\nabla m_k(\theta)$ is a descent direction of $f(\theta)$, and to search for a suitable optimizer gain $\mu\in\mathbb{R}_{> 0}$ for which the Armijo condition \eqref{eq:Armijo} is satisfied. We perform both these steps without requiring additional measurements to be performed, by studying the bounds $\underaccent{\bar}{b}(\theta)$ and $\bar{b}(\theta,\mu)$ in \eqref{eq:nabla_m_descent_bound} and \eqref{eq:Armijo_bound}. To construct expressions for these bounds, we define the symmetric matrix
\begin{align}
    \mathfrak{K}(\theta) := \begin{bmatrix}
        K_{\Omega\Omega} & \nabla K_\Omega(\theta)^\intercal\\
        \nabla K_\Omega(\theta) & D^2_{1,2}\kappa(\theta,\theta)
    \end{bmatrix}\label{eq:K_theta}
\end{align}
with $K_{\Omega\Omega}$ the $N_k \times N_k$ matrix whose $(i,j)$-th element is given by $\kappa(\widetilde{\theta}_i, \widetilde{\theta}_j)$ (cf. Definition~\ref{def:positive_definite_kernel}), $\nabla K_\Omega(\theta)$ a matrix of partial derivatives of the kernel functions as in \eqref{eq:nabla_K_Omega}, and 
\begin{align}
        D^2_{1,2}&\kappa(\theta,\theta) :=\nonumber\\
        &\begin{bmatrix} D^{(e_1,e_1)}\kappa(\theta, \theta) & \cdots & D^{(e_1,e_{n_\theta})}\kappa(\theta,\theta)\\
        \vdots & \ddots & \vdots\\
        D^{(e_{n_\theta}, e_1)}\kappa(\theta,\theta) & \cdots & D^{(e_{n_\theta},e_{n_\theta})}\kappa(\theta,\theta)
        \end{bmatrix}
\end{align}
a matrix of second-order partial derivatives with $D^{(e_i,e_j)} \kappa(\theta, \theta)$ as in \eqref{eq:kernel_derivatives}. Moreover, we define the symmetric matrix
\begin{align}
    \mathfrak{K}^\prime&(\theta,\theta^+) :=  \nonumber \\
    &\begin{bmatrix}
        K_{\Omega\Omega} & K_\Omega(\theta)^\intercal & K_\Omega(\theta^+)^\intercal & \nabla K_{\Omega}(\theta)^\intercal\\
        K_\Omega(\theta) & \kappa(\theta,\theta) &  \kappa(\theta,\theta^+) & \nabla K(\theta, \theta)^\intercal\\
        K_\Omega(\theta^+) & \kappa(\theta^+,\theta) & \kappa(\theta^+,\theta^+) & \nabla K(\theta, \theta^+)^\intercal\\
        \nabla K_{\Omega}(\theta) & \nabla K(\theta, \theta) & \nabla K(\theta, \theta^+) &D^2_{1,2}\kappa(\theta,\theta)
    \end{bmatrix}\label{eq:K_prime_theta}
\end{align}
with $K_\Omega(\cdot)$ a row vector of kernel functions centered at the inputs $\Omega = \{\widetilde{\theta}_1, \widetilde{\theta}_2, \dots, \widetilde{\theta}_{N_k}\}$ as in \eqref{eq:m_k_sum}, and where
\begin{equation} 
    \nabla K(a,b) := \begin{bmatrix} D^{(e_1,0)}\kappa(a,b) & \cdots & D^{(e_{n_\theta},0)}\kappa(a,b) \end{bmatrix}^\intercal. \label{eq:nabla_K_ab}
\end{equation}
Note that $\mathfrak{K}^\prime(\theta, \theta^+)$ in \eqref{eq:K_prime_theta} is what we obtain if $\mathfrak{K}(\theta)$ in \eqref{eq:K_theta} gets appended with evaluations of the kernel (derivatives) at $\theta$ and $\theta^+$, i.e., if $\Omega$ in \eqref{eq:K_theta} gets replaced by $\Omega^\prime := \Omega \cup \{\theta,\theta^+\}$.

Given the definitions \eqref{eq:K_theta} and \eqref{eq:K_prime_theta}, the following theorem states that each bound $\underaccent{\bar}{b}(\theta)$ and $\bar{b}(\theta, \mu)$ can be evaluated by solving a second-order cone program (SOCP) for a given input $\theta \in \mathbb{R}^{n_\theta}$ and optimizer gain $\mu \in \mathbb{R}_{>0}$.
\begin{thm}
    \label{thm:m_sufficiently_accurate}
    Suppose Assumptions~\ref{as:bounded_measurment_error} and \ref{as:true_function_in_RKHS} hold. Let a dataset $\mathcal{D}_k = \{(\widetilde{\theta}_i, \widetilde{y}_i) \mid i = 1,2,\dots, N_k \}$ of constant inputs $\widetilde{\theta}_i$ and corresponding output measurements $\widetilde{y}_i$, e.g., as in \eqref{eq:theta_tilde} and \eqref{eq:sampler_y}, respectively, be given. Furthermore, let $m_k(\cdot)$ be an approximation of the input-output map $f(\cdot)$ on the basis of this dataset $\mathcal{D}_k$ as in \eqref{eq:m_k_eSVR}. Then the minimum value that $\nabla f(\theta)^\intercal \nabla m_k(\theta)$ can attain at a given input $\theta \in \mathbb{R}^{n_\theta}$ can be found by solving the SOCP
    \begin{subequations}
        \label{eq:b_nabla_upper}
        \begin{align}
            \underaccent{\bar}{b}(\theta) = & \nonumber \\
            \underset{\beta \in \mathbb{R}^{N}}{\min} \: & \begin{bmatrix} 0_{1\times N_k} & \nabla m_k(\theta)^\intercal \end{bmatrix}\mathfrak{K}(\theta)\beta\\
            {\rm s.t.} \: & \beta^\intercal \mathfrak{K}(\theta) \beta \leq \Gamma^2\\
            & \lvert \xi_i^\intercal \mathfrak{K}(\theta) \beta - \widetilde{y}_i \rvert \leq \bar{\delta} \quad \forall i \in\{1,2,\dots,N_k\},\label{eq:perturbation_bound_constraint_nabla}
        \end{align}
    \end{subequations}  
    with $N = N_k + n_\theta$, $\mathfrak{K}(\theta)$ as in \eqref{eq:K_theta} and $\xi_i \in \mathbb{R}^{N}$ a vector with $i$-th element equal to one and other elements equal to zero.

    Moreover, for any given $\mu \in \mathbb{R}_{>0}$ and $c \in (0,1)$, the maximum value that $f(\theta^+) - f(\theta) + c \mu \nabla f(\theta)^\intercal \nabla m_k(\theta)$ with $\theta^+$ as in \eqref{eq:theta_plus_kb} can attain at a given input $\theta \in \mathbb{R}^{n_\theta}$ can be found by solving the SOCP
    \begin{subequations}
        \label{eq:b_A_upper}
        \begin{align}
            \bar{b}(\theta, &\mu) = \nonumber \\
            \underset{\beta \in \mathbb{R}^{N}}{{\rm max}} \: & \begin{bmatrix} 0_{1\times N_k}  & -1 & 1 & c \mu \nabla m_k(\theta)^\intercal \end{bmatrix}\mathfrak{K}^\prime(\theta,\theta^+) \beta\\
            {\rm s.t.} \: & \beta^\intercal \mathfrak{K}^\prime(\theta,\theta^+)\beta \leq \Gamma^2\\
            & \lvert \xi_i^\intercal \mathfrak{K}^\prime(\theta,\theta^+) \beta - \widetilde{y}_i \rvert \leq \bar{\delta} \; \forall i \in \{1,2,\dots, N_k\}\label{eq:perturbation_bound_constraint}
        \end{align}
    \end{subequations} 
    with $N = N_k + 2 + n_\theta$, $\mathfrak{K}^\prime(\theta,\theta^+)$ as in \eqref{eq:K_prime_theta}, and $\xi_i \in \mathbb{R}^{N}$ again a vector with $i$-th element equal to one and other elements equal to zero. 
\end{thm}
\begin{pf}
    The proof of Theorem~\ref{thm:m_sufficiently_accurate} can be found in Appendix~\ref{app:proof_m_sufficiently_accurate}.  \hfill $\square$
\end{pf}
\begin{rem}
    The inputs $\Theta = [\widetilde{\theta}_1 \,|\, \cdots \,|\, \widetilde{\theta}_{N_k} \,|\, \theta \,|\, \theta^+]$ used in \eqref{eq:b_nabla_upper} and \eqref{eq:b_A_upper} might not be pairwise distinct, potentially causing the solutions to the SOCPs to be non-unique. Therefore, in case $\Theta_i=\Theta_j$, $i\neq j$, with $\Theta_i$ and $\Theta_j$ denoting the $i$-th and $j$-th column  of $\Theta$, respectively, one removes the rows and columns corresponding to $\Theta_i$ from \eqref{eq:b_nabla_upper} and/or \eqref{eq:b_A_upper} to maintain uniqueness of the solution. This follows from the observation that in the proof of Theorem~\ref{thm:m_sufficiently_accurate} removing $\kappa(\cdot,\Theta_i)$ from the span in \eqref{eq:H_parallel} does not change $\mathcal{H}^\parallel$ (the finite-dimensional subspace of $\mathcal{H}$ spanned by the kernel slices and their partial derivatives evaluated at the inputs in $\Theta$).
\end{rem}
With the bounds $\underaccent{\bar}{b}(\theta)$ and $\bar{b}(\theta,\mu)$ as in Theorem~\ref{thm:m_sufficiently_accurate}, we state in Algorithm~\ref{alg:kb_esc} the proposed kernel-based extremum-seeking control algorithm that we outlined briefly in Section~\ref{sec:kernel-based_extremum-seeking} (see also Figure~\ref{fig:block_scheme}). In this algorithm, we perform the backtracking line search over $\mu \in [\underaccent{\bar}{\mu}, \bar{\mu}]$ with $\underaccent{\bar}{\mu}$ and  $\bar{\mu}$ a positive lower and upper bound, respectively, by multiplying $\mu$ by a reduction factor $\varrho \in (0,1)$ until the Armijo condition \eqref{eq:Armijo} is guaranteed to be satisfied or $\mu < \underaccent{\bar}{\mu}$.  
\begin{algorithm}
    \caption{Kernel-based extremum-seeking control}
    \label{alg:kb_esc}
    \begin{algorithmic}[1]
        \small
    \Require Waiting time $T$ and number of measurements $n_v$ for a single standard parameter update step as in \eqref{eq:theta_tilde}-\eqref{eq:Ytilde}; upper bound $\bar{\delta}$ on the measurement perturbations as in Assumption~\ref{as:bounded_measurment_error}; upper bound $\Gamma$ on $\lVert f \rVert$ as in Assumption~\ref{as:true_function_in_RKHS}; backtracking line search lower bound $\underaccent{\bar}{\mu}$, upper bound $\bar{\mu}$ and reduction factor $\varrho\in(0,1)$; control parameter $c \in (0,1)$ as in the Armijo condition \eqref{eq:Armijo}; maximum number of update steps $\bar{k}$; initial optimization algorithm state $\widehat{\theta}_0$; and a dataset $\mathcal{D}_0$ that is either empty, or contains previously obtained input-output data pairs $(\widetilde{\theta}, \widetilde{y})$, e.g., as in \eqref{eq:theta_tilde}-\eqref{eq:sampler_y}.
    \Ensure Found optimal parameters $\widehat{\theta}_{\bar{k}}$.
    \State{Initialize the algorithm update index $k$, i.e, $k \gets 0$.}
    \If{$\mathcal{D}_0$ is empty}
        \State{\parbox[t]{\dimexpr\linewidth-1.5em}{Perform a standard parameter update step as in \eqref{eq:theta_tilde}-\eqref{eq:Ytilde} to update the optimizer state $\widehat{\theta}_0$.\strut}\label{line:standard_update_start}}
        \State{\parbox[t]{\dimexpr\linewidth-1.5em}{Add data collected during the update step to a dataset $\mathcal{D}_1$, i.e., $\mathcal{D}_1 \gets \{(\widetilde{\theta}_i, \widetilde{y}_i) \mid i = 1, 2, \dots, n_v\}$.\strut}}
        \State{Update the algorithm update index $k$, i.e, $k \gets 1$.}
    \EndIf
    \While{$k < \bar{k}$}
        \State{\parbox[t]{\dimexpr\linewidth-1.5em}{Construct an approximation $m_k(\cdot)$ of the input-output map $f(\cdot)$ from $\mathcal{D}_k$ as in \eqref{eq:m_k_eSVR}.\strut}}
        \State{\parbox[t]{\dimexpr\linewidth-1.5em}{\textbf{if} $-\nabla m_k(\widehat{\theta}_k)$ is guaranteed to be a descent direction of $f(\cdot)$, i.e., if $\underaccent{\bar}{b}(\widehat{\theta}_k) > 0$ with $\underaccent{\bar}{b}(\cdot)$ as in \eqref{eq:b_nabla_upper}  (cf. \eqref{eq:nabla_m_descent_bound}) \textbf{then}\strut}}
            \State{\hspace{1.5em} \parbox[t]{\dimexpr\linewidth-3em-\fboxsep}{Perform a backtracking line search over $\mu\in[\underaccent{\bar}{\mu}, \bar{\mu}]$, i.e., set $\mu \gets \bar{\mu}$ and\strut}}
            \State{\hspace{1.5em} \parbox[t]{\dimexpr\linewidth-3em-\fboxsep}{\textbf{while} $\mu \geq \underaccent{\bar}{\mu}$ \textbf{and} $\bar{b}(\widehat{\theta}_k, \mu) > 0$ with $\bar{b}(\cdot,\mu)$ as in \eqref{eq:b_A_upper}\strut}}
                    \State{\hspace{3em} \parbox[t]{\dimexpr \linewidth-4.5em-\fboxsep}{Reduce $\mu$, i.e., $\mu \gets \varrho \mu$.\strut}}
                    \State{\hspace{1.5em} \textbf{end while}}
                    \State{\hspace{1.5em} \parbox[t]{\dimexpr \linewidth-3em-\fboxsep}{\textbf{if} the Armijo condition \eqref{eq:Armijo} is guaranteed to be satisfied, i.e., if  $\bar{b}(\widehat{\theta}_k,\mu) \leq 0$ (cf. \eqref{eq:Armijo_bound}) \textbf{then}\strut}}
                        \State{\hspace{3em} \parbox[t]{\dimexpr \linewidth-4.5em-\fboxsep}{Perform a kernel-based update \eqref{eq:theta_plus_kb} without performing additional measurements, i.e., $\widehat{\theta}_{k+1} \gets \widehat{\theta}_k - \mu \nabla m_k(\widehat{\theta}_k)$, $\mathcal{D}_{k+1} \gets \mathcal{D}_k$, and $N_{k+1} \gets N_k$.\strut}\label{line:kb_update}}
                        \State{\hspace{1.5em} \parbox[t]{\dimexpr \linewidth-3em-\fboxsep}{\textbf{else}\strut}}
                        \State{\hspace{3em} \parbox[t]{\dimexpr \linewidth-4.5em-\fboxsep}{Jump to Line \ref{line:standard_update}.\strut}}
                    \State{\hspace{1.5em} \textbf{end if}}
                    \State{\textbf{else}}
                        \State{\hspace{1.5em} \parbox[t]{\dimexpr\linewidth-3em-\fboxsep}{Perform a standard parameter update step as in \eqref{eq:theta_tilde}-\eqref{eq:Ytilde} to update the optimizer state $\widehat{\theta}_k$.\strut}}\label{line:standard_update}
                        \State{\hspace{1.5em} \parbox[t]{\dimexpr\linewidth-3em-\fboxsep}{Add data collected during the update step to a dataset $\mathcal{D}_{k+1}$, i.e., $\mathcal{D}_{k+1} \gets \mathcal{D}_k \cup \{(\widetilde{\theta}_i, \widetilde{y}_i) \mid i = N_k + 1, \dots, N_k + n_v\}$ and $N_{k+1} \gets N_k + n_v$.\strut}}
                    \State{\textbf{end if}}
                \State{\parbox[t]{\dimexpr\linewidth-1.5em-\fboxsep}{Update the algorithm update index, i.e., $k \gets k + 1$.\strut}}
    \EndWhile
    \State{\Return $\widehat{\theta}_{\bar{k}}$.}
    \end{algorithmic}
\end{algorithm}

\section{Stability analysis for kernel-based extremum-seeking control}
\label{sec:stability_analysis}

In this section, we perform a stability analysis for the kernel-based extremum-seeking approach described in Algorithm~\ref{alg:kb_esc}. To this end, we define the function (see also \citet[Equation~(22)]{Hazeleger2022})
\begin{align}
    W(\psi, x, \theta) := V_{p}(\psi, x) + 2V(\theta), \label{eq:definition_W}
\end{align}
where $V(\cdot)$ is as in \eqref{eq:definition_V} and
\begin{align}
    V_{p}(\psi, x) := \lVert x \rVert_{\mathcal{A}(\psi + v_{n_v}(\psi))} + \eta \lVert \psi \rVert_{\mathcal{C}} \label{eq:definition_Vp}
\end{align}
with $\eta \in \mathbb{R}_{> 0}$ a constant, $x$ the system state, and $\psi$ a memory state for the optimizer state at the start of the last standard update step. That is, $\widetilde{\theta}_{N_k} = \psi + v_{n_v}(\psi)$ is the last input that has been applied to the system (cf. \eqref{eq:theta_tilde} and Lines~\ref{line:standard_update_start} and \ref{line:standard_update} of Algorithm~\ref{alg:kb_esc}). Note that in \eqref{eq:definition_W}, $V_p(\psi, x)$ contains terms relating to the distance of the system state to the attractor $\mathcal{A}(\psi + v_{n_v}(\psi))$ (cf. Assumption~\ref{as:global_attractor}) corresponding to the last applied input $\psi + v_{n_v}(\psi)$, and the distance from the corresponding optimizer state $\psi$ to the set of minimizers $\mathcal{C}$ (cf. \eqref{eq:definition_Vp}). That is, $V_p(\psi, x)$ is small if the system state $x$ is close to the attractor corresponding to the last applied input and the corresponding optimizer state $\psi$ is close to the set of minimizers $\mathcal{C}$, whereas $V(\theta)$ in \eqref{eq:definition_W} relates to the distance of the current optimizer state $\theta$ to the set of minimizers $\mathcal{C}$ since $\omega_1(\lVert \theta \rVert_\mathcal{C}) \leq V(\theta)\leq \omega_2(\lVert \theta \rVert_\mathcal{C})$ (cf. Assumption~\ref{as:DV_standard}).  Thus, $W(\psi, x, \theta)$ is small when both the current optimizer state $\theta$ and memory state $\psi$ are close to $\mathcal{C}$, and $x$ is close to the attractor corresponding to the last applied input $\psi + v_{n_v}(\psi)$. 

The following two lemmas state that there exist upper bounds on the increment
\begin{align}
    \Delta W(\psi, x, \theta) & := W(\psi^+, x^+, \theta^+) - W(\psi, x, \theta) \label{eq:definition_DW}
\end{align}
during standard update steps, and the increment
\begin{align}
    \Delta V(\theta) & := V(\theta^+) - V(\theta) \label{eq:definition_DV}
\end{align}
during kernel-based update steps, respectively, which will prove to be useful in the stability analysis.
\begin{lem}
    \label{lem:upper_bound_DW_standard}
    Suppose that the system $\Sigma_p$ in Definition~\ref{def:Sigma_p} satisfies Assumption~\ref{as:Sigma_p}, and that the optimization algorithm $\Sigma$ in \eqref{eq:Sigma} satisfies Assumption~\ref{as:Sigma}. Let $x$ and $\theta$ denote, respectively, the state of $\Sigma_p$ and $\Sigma$ at the start of the current algorithm update step $k$. Furthermore, let $\psi$ denote a memory state for optimizer state at the start of the last standard update step, i.e., $\widetilde{\theta}_{N_k} = \psi + v_{n_v}(\psi)$ is the last input that has been applied to the system (cf. \eqref{eq:theta_tilde}). Then, for any $\Delta_\theta, \Delta_x \in \mathbb{R}_{> 0}$ such that $\lVert \theta \rVert_{\mathcal{C}}  \leq \Delta_\theta$, $\lVert \psi \rVert \leq \Delta_\theta$, and $\lVert x \rVert_{\mathcal{A}(\psi + v_{n_v}(\psi))} \leq \Delta_x$, there exists a sufficiently long waiting time $T^*\geq T$, with $T$ as in Assumption~\ref{as:sufficiently_long_T}, and a class-$\mathcal{K}_\infty$ function $\widetilde{\gamma}(\cdot)$, such that for a standard parameter update step as in \eqref{eq:theta_tilde}-\eqref{eq:Ytilde} the upper bound
    \begin{align}
        \Delta W(\psi, x, \theta) & \leq -\widetilde{\gamma}\left(W(\psi, x, \theta)\right) + \gamma + 2\delta + \delta_V\label{eq:bound_DW_standard}
    \end{align}
    with $\Delta W(\psi, x, \theta)$ as in \eqref{eq:definition_DW}, holds for arbitrarily small $\delta, \gamma, \delta_V \in \mathbb{R}_{>0}$.
\end{lem}
\begin{pf}
    Note that Assumptions~\ref{as:Sigma_p} and \ref{as:Sigma} imply satisfaction of \citet[Assumptions 2, 3, 7, and 11]{Hazeleger2022} (with number of constraints equal to zero). Hence, Lemma~\ref{lem:upper_bound_DW_standard} follows directly from the increment in \citet[Theorem~13, Equation (24)]{Hazeleger2022} by taking the number of constraints equal to zero. \hfill $\square$
\end{pf}

\begin{lem}
    \label{lem:upper_bound_DV_kernel-based}
    Suppose Assumptions~\ref{as:minimum_f}, \ref{as:DV_standard}, \ref{as:bounded_measurment_error} and \ref{as:true_function_in_RKHS} hold. Let $m_k(\cdot)$ be an approximation of $f(\cdot)$ obtained as in \eqref{eq:m_k_eSVR} from a dataset $\mathcal{D}_k = \{ \widetilde{\theta}_i, \widetilde{y}_i \mid i = 1,2, \dots, N_k\}$ with constant inputs $\widetilde{\theta}_i$ and their corresponding output measurements $\widetilde{y}_i$, e.g., as in \eqref{eq:theta_tilde} and \eqref{eq:sampler_y}, respectively. Furthermore, let $\theta^+$ be the result of a kernel-based update step as in \eqref{eq:theta_plus_kb}. Finally, let $\underaccent{\bar}{b}(\theta)$ and $\bar{b}(\theta, \mu)$ be bounds as in Theorem~\ref{thm:m_sufficiently_accurate}. If, for any $\theta \in \mathbb{R}^{n_\theta} \backslash \mathcal{C}$, $\underaccent{\bar}{b}(\theta) > 0$ and $\mu \in \mathbb{R}_{> 0}$ in \eqref{eq:theta_plus_kb} is chosen such that $\bar{b}(\theta, \mu) \leq 0$, then for any $\Delta_\theta \in \mathbb{R}_{> 0}$ such that $\lVert \theta \rVert_{\mathcal{C}} \leq \Delta_\theta$, there exists a class-$\mathcal{K}$ function $\widetilde{\rho}_k(\cdot)$ such that
    \begin{align}
        \Delta V(\theta) & \leq -\widetilde{\rho}_k \left( V(\theta) \right)\label{eq:bound_DV_kernel-based}
    \end{align}
    with $\Delta V(\theta)$ as in \eqref{eq:definition_DV}.
\end{lem}
\begin{pf}
    The proof of Lemma~\ref{lem:upper_bound_DV_kernel-based} can be found in Appendix~\ref{app:proof_upper_bound_DV_kernel-based}. \hfill $\square$
\end{pf}
\begin{rem}
Note with respect to Remark~\ref{rm:assumptions_optimizer} and Lemmas~\ref{lem:upper_bound_DW_standard} and \ref{lem:upper_bound_DV_kernel-based}, that when $V(\cdot)$ would be chosen as general locally Lipschitz function, instead of as in \eqref{eq:definition_V}, it is not immediately clear under which conditions a kernel-based update step is `helpful', i.e., under which conditions a kernel-based update step leads to a decrease in $V(\theta)$. By defining $V(\cdot)$ explicitly as in \eqref{eq:definition_V}, such decrease can be guaranteed by satisfaction of the Armijo condition \eqref{eq:Armijo} if $-\nabla m_k(\theta)$ is a descent direction as shown in the proof of Lemma~\ref{lem:upper_bound_DV_kernel-based}.
\end{rem}

The next theorem combines the results from Lemmas~\ref{lem:upper_bound_DW_standard} and \ref{lem:upper_bound_DV_kernel-based}, and states the requirements on the initial conditions and the parameters of the kernel-based extremum-seeking control algorithm described in Algorithm~\ref{alg:kb_esc}, such that the optimizer state $\theta$ converges to an arbitrarily small neighborhood of the set of minimizers $\mathcal{C}$.
\begin{thm}
    \label{thm:stability_kernel-based_ESC}
    Suppose that the system $\Sigma_p$ in Definition~\ref{def:Sigma_p} and optimizer $\Sigma$ in \eqref{eq:Sigma} satisfy Assumptions~\ref{as:Sigma_p}, \ref{as:Sigma}, \ref{as:bounded_measurment_error}, and \ref{as:true_function_in_RKHS}. For any $\theta_0, \psi_0\in\mathbb{R}^{n_\theta}$ and $x_0 \in \mathcal{X}$ with $\lVert \theta_0 \rVert_{\mathcal{C}} \leq \Delta_\theta$, $\lVert \psi_0 \rVert_{\mathcal{C}} \leq \Delta_\theta$ and $\lVert x_0 \rVert_{\mathcal{A}(\psi_0+v_{n_v}(\psi_0))} \leq \Delta_x$ for some $\Delta_\theta,\Delta_x\in\mathbb{R}_{>0}$, there exists a sufficiently long waiting time $T^*\geq T$ for standard parameter updates as in \eqref{eq:theta_tilde}-\eqref{eq:Ytilde}, with $T$ satisfying Assumption~\ref{as:sufficiently_long_T}, and a class-$\mathcal{K}_\infty$ function $\varphi(\cdot)$, such that for the kernel-based extremum-seeking algorithm described in Algorithm~\ref{alg:kb_esc} the optimizer state $\theta$ converges to the set
    \begin{align}
        \left\{\theta \in \mathbb{R}^{n_\theta} \mid \lVert \theta \rVert_{\mathcal{C}} \leq \omega_1^{-1} \left( \frac{1}{2} \varphi( \gamma + 2\delta + \delta_V)\right)\right\}
    \end{align}
    as $k \to \infty$ with $\omega_1 \in \mathcal{K}_\infty$ as in Assumption~\ref{as:DV_standard}, and where $\gamma$, $\delta$, and $\delta_V$ can be made arbitrarily small.
\end{thm}
\begin{pf}
    The proof of Theorem~\ref{thm:stability_kernel-based_ESC} can be found in Appendix~\ref{app:proof_stability_kernel-based_ESC}.  \hfill $\square$
\end{pf}
\begin{rem}
    Note that the input applied to the system $\Sigma_p$ is not changed during kernel-based update steps (cf. Algorithm~\ref{alg:kb_esc}, Line~\ref{line:kb_update}), and that the duration of kernel-based update steps is negligible compared to the timescale at which the system dynamics evolve (since there is no waiting time $T$ for kernel-based update steps) and can therefore be considered instantaneous. As a consequence, the memory state $\psi$ and system state $x$ do not change (significantly) during kernel-based update steps. Therefore, while the optimizer state $\theta$ converges to a neighborhood of the set of minimizers as stated in Theorem~\ref{thm:stability_kernel-based_ESC}, the system output $y$ does not necessarily converge of a neighborhood of the minimum of the input-output map $f(\cdot)$. However, by Assumptions~\ref{as:global_attractor} and \ref{as:f_well-defined}, this discrepancy between the value of the steady-state input-output map associated with the current optimizer state and the real plant output can always be made arbitrarily small by applying the optimizer state as an input to the plant and waiting sufficiently long (after the last kernel-based update step).  
\end{rem}
\section{Simulation study}
\label{sec:simulation_study}
In this section, we show the benefits of kernel-based extremum-seeking control in a simulation example and compare its performance to the standard sampled-data approach as in, e.g., \cite{Teel2001} and \cite{Hazeleger2022}. To this end, we consider the nonlinear multi-input-single-output dynamical system
\begin{subequations}
    \label{eq:simulation_system}
    \begin{align}
        \dot{x}_1(t) &= -4x_1(t)^3 +0.5(\theta_1(t) - \theta_1^*)^6,\\
        \dot{x}_2(t) &= 2x_1(t) - 5x_2(t)^3 + (\theta_2(t) - \theta_2^*)^2,\\
        y(t) &= -\exp{(-0.1x_2(t)^3)}
    \end{align} 
\end{subequations}
with $\theta^* = [\begin{array}{@{}cc@{}}\theta_1^* & \theta_2^*\end{array}]^\intercal$. For each standard update step (in either approach), we use a gradient-based update step, where the gradient of the input-output map is estimated using the central difference scheme. To this end, we perform $n_v=4$ measurements per update step, during which constant dither $v_j$, $j=1, 2,\dots, n_v$, with amplitude $c_v$ is added to the current optimizer state $\widehat{\theta}_k$ along the positive and negative axes of the parameters $\theta_1$ and $\theta_2$, respectively, to obtain the inputs $\widetilde{\theta}_{N_k+j}$ as in \eqref{eq:theta_tilde}, i.e., $v_1$ to $v_4$ are given as
\begin{align*}
    \left[
    \begin{array}{@{}c|c|c|c@{}}
         v_1 & v_2 & v_3 & v_4
    \end{array}\right] =
    \left[
    \begin{array}{@{}c|c|c|c@{}}
        -c_v & c_v & 0 & 0\\ 
        0 & 0 & -c_v & c_v
    \end{array}\right].
\end{align*} 
We use the corresponding output measurements $\widetilde{y}_{N_k+j}$ (cf. \eqref{eq:sampler_y}), in an update step of the form
\begin{align}
    \widehat{\theta}_{k+1} = \widehat{\theta}_{k} - \frac{\widetilde{\mu}}{2c_v}\begin{bmatrix}
        \widetilde{y}_{N_k+2}-\widetilde{y}_{N_k+1}\\
        \widetilde{y}_{N_k+4}-\widetilde{y}_{N_k+3}
    \end{bmatrix}
\end{align}
with fixed optimizer gain $\widetilde{\mu}$. To construct the approximation $m_k(\theta)$ as in \eqref{eq:m_k_eSVR}, we use the squared exponential kernel $\kappa(\theta,\theta^\prime)$ as in \eqref{eq:SE_kernel} with length scale $\sigma = 5$. Note that for the system \eqref{eq:simulation_system} the unknown steady-state input-output map $f(\theta)$ is indeed given by $f(\theta) = -\kappa(\theta,\theta^*)$. Furthermore, we choose the upper bound $\Gamma$ from Assumption~\ref{as:true_function_in_RKHS} to be 5\% higher than the true norm of $f$ (which is 1). All other used parameter values are listed in Table~\ref{tab:parameter_values}.
\begin{table}[tb]
    \centering
    \caption{Parameter values used in the simulation example.}
    \label{tab:parameter_values}
    \begin{tabular}{*{5}{|c}|}
        \hline
        $\theta^*$  & $c_v$&  $\widetilde{\mu}$ & $T$ & $\bar{\delta}$ \\
        \hline
        $[\begin{array}{@{}cc@{}}3 & 1\end{array}]^\intercal$ & $10^{-2}$  & $5$ & $10$ & $2.5\cdot10^{-3}$\\
        \hline\hline
        $\Gamma$ & $c$ & $\underaccent{\bar}{\mu}$ & $\bar{\mu}$ & $\varrho$ \\
        \hline
        $1.05$ & $10^{-4}$ & $0.1$ & $50$ & $0.9$\\
        \hline
    \end{tabular} 
\end{table}

\begin{figure}
    \centering
    \includegraphics[width=\linewidth]{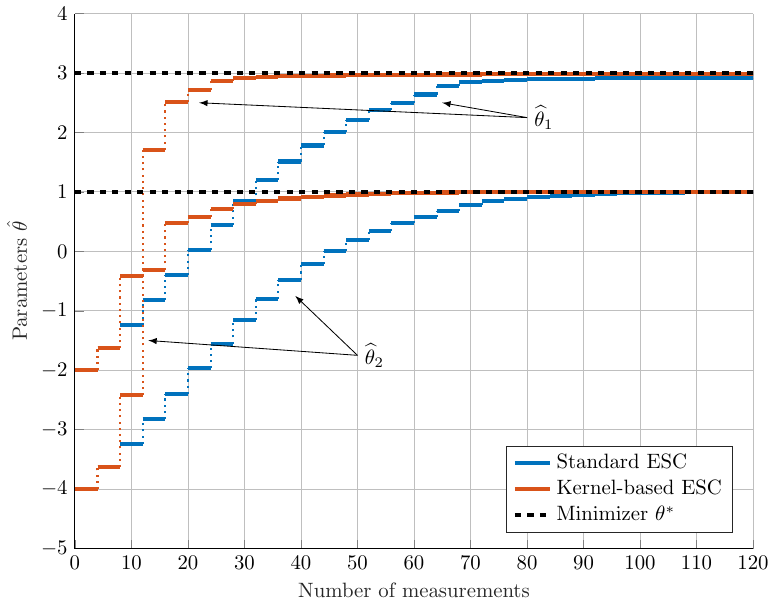}
    \caption{The optimizer state $\widehat{\theta}$ for both ESC approaches as a function of the number of measurements. The proposed kernel-based approach requires only 60 measurements to reach the same neighborhood of the optimum that the standard approach reaches with 100 measurements.}
    \label{fig:thetahat_nr_exp}
\end{figure}

\begin{figure}
    \centering
    \includegraphics[width=\linewidth]{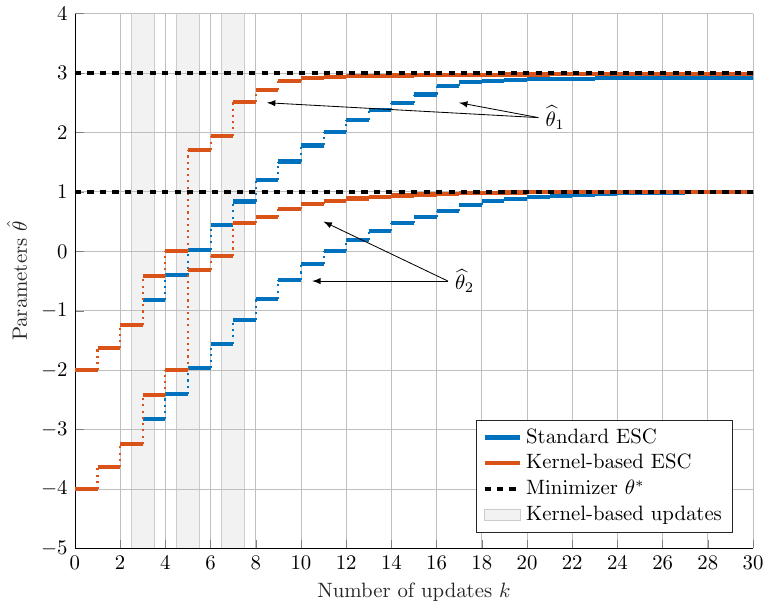}
    \caption{The optimizer state $\widehat{\theta}$ for both ESC approaches as a function of the number of update steps. The third, fifth, and seventh update in the proposed kernel-based approach are kernel-based update steps, which do not require additional measurements to be performed to perform the update and allow searching for a suitable optimizer gain, resulting in needing only 18 update steps to reach the same neighborhood of the optimum as the standard approach does with 25 update steps.}
    \label{fig:thetahat_nr_updates}
\end{figure}

Starting both approaches with initial conditions $x(0) = [\begin{array}{@{}cc@{}}0 & 0\end{array}]^\intercal$ and $\widehat{\theta}_0 =[\begin{array}{@{}cc@{}}-2 &-4\end{array}]^\intercal$, shows that using the kernel-based approach the optimizer state $\widehat{\theta}$ converges with 60 measurements to the same neighborhood of the minimizer $\theta^*$ as the standard approach reaches with 100 measurements, as illustrated by Figure~\ref{fig:thetahat_nr_exp}. The number of measurements needed to reach a small neighborhood of the minimizer is thus reduced by 40\%. This reduction is the result of two benefits that the kernel-based approach provides. First, three of the update steps performed in the kernel-based approach ($k=3, 5$, and $7$) are kernel-based update steps, as illustrated in Figure~\ref{fig:thetahat_nr_updates}, which do not require additional measurements to be performed to update the parameters. Second, these kernel-based update steps are larger due to the use of a larger optimizer gain ($\mu=14.1$, $26.6$, and $17.4$, respectively) computed by the backtracking line search, compared to the fixed value used in the standard update steps ($\widetilde{\mu} = 5$). Because of these larger steps, the optimizer state $\widehat{\theta}$ converges in fewer update steps to the same neighborhood of the minimizer $\theta^*$ than the standard approach: 18 instead of 25 updates, which is a reduction of 28\%.

\begin{figure}
    \centering
    \includegraphics[width=\linewidth]{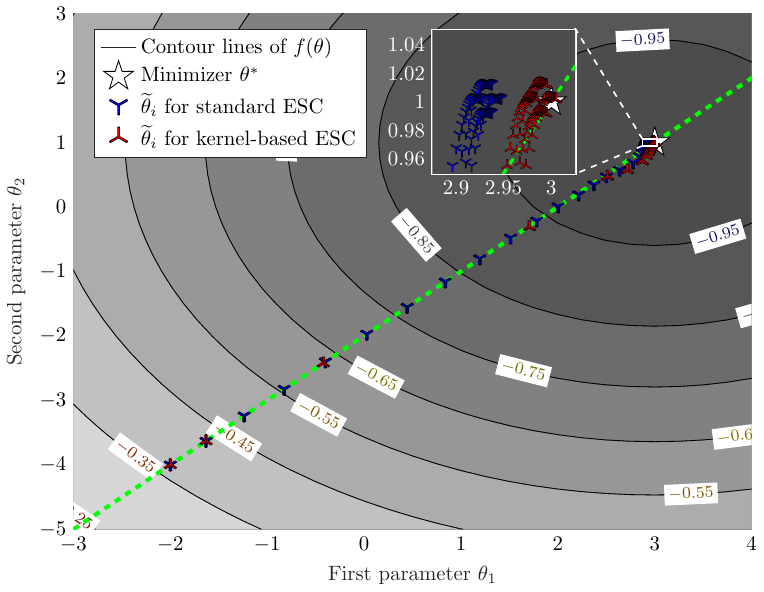}
    \caption{Contour plot of the steady-state input-output map $f(\theta)$, along with inputs $\widetilde{\theta}_i$ applied during measurements in both approaches. These applied inputs lie close to the line passing through the initial optimizer state $\widehat{\theta}_0$ and the minimizer $\theta^*$ (shown in green).}
    \label{fig:contour}
\end{figure}

\begin{figure}
    \centering
    \includegraphics[width=\linewidth]{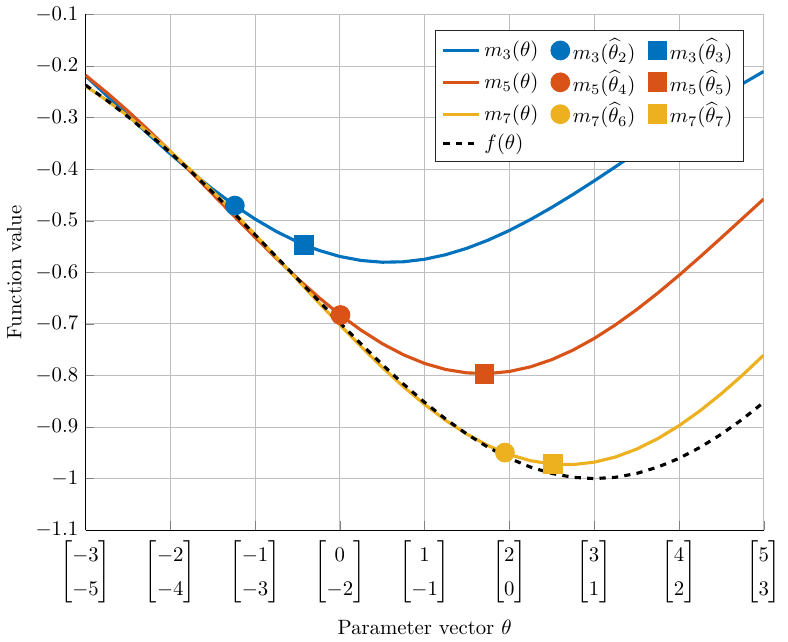}
    \caption{Cross-section of the input-output map $f(\theta)$ as well as the approximation $m_k(\theta)$ at different kernel-based update steps, along the search direction (cf. green line in Figure~\ref{fig:contour}). $m_k(\theta)$ quickly improves, allowing large kernel-based update steps to be taken from the circle to the square.}
    \label{fig:function_values}
\end{figure}

Performing these large kernel-based steps is possible because by using the input-output data collected during regular operation of the extremum-seeking controller to construct and update the approximation $m_k(\theta)$ online, it quickly becomes an accurate representation of the true input-output map $f(\theta)$ along the search direction. To illustrate this, we note that most of the inputs $\widetilde{\theta}_i$ applied during the measurements in both approaches lie close to the line through the initial optimizer state $\widehat{\theta}_0$ and the minimizer $\theta^*$, as shown in Figure~\ref{fig:contour}. Plotting the cross-section of the input-output map $f(\theta)$ and the approximation $m_k(\theta)$ at the three kernel-based update steps ($k=3, 5$ and $7$) along this line indeed shows that along this search direction the approximation $m_k(\theta)$ quickly improves, as shown in Figure~\ref{fig:function_values}.

\section{Conclusion}
\label{sec:conclusion}

Motivated by the potentially costly and time-consuming nature of performing measurements in practical applications, we presented a novel extremum-seeking approach aimed at reducing the total number of measurements needed to optimize performance. The proposed approach uses kernel-based function approximation to construct online an approximation of the steady-state input-output map of the system, based on data collected during regular extremum-seeking steps. We use this approximation to determine a suitable search direction and optimizer gain to perform a parameter update without performing additional measurements whenever it is sufficiently accurate to guarantee a decrease in the input-output map. If such a decrease cannot be guaranteed, we perform a regular extremum-seeking parameter update step to update the parameters. In addition, we use the data collected during this update step to improve the approximation for the next parameter update step. By using this approach, reductions in both the number of performed measurements and the number of update steps can be obtained, as illustrated by a simulation study in which the number of measurements and updates are reduced by 40\% and 28\%, respectively.

Future research could consider extensions to a more general class of optimization methods as in \cite{Khong2013a,Khong2013b}, for example, for both the standard parameter update steps and kernel-based update steps. Other potential research directions include the use of variable waiting times instead of a fixed one, as, e.g., in \citet{Poveda2017}, to reduce the time-consuming nature of performing measurements even further, or extensions in which only a subset of all collected data are used, e.g., to reduce computational complexity or to deal with input-output maps that are slowly changing over time.

\appendix
\section{Proof of Theorem~\ref{thm:m_sufficiently_accurate}}
\label{app:proof_m_sufficiently_accurate}

We start the proof by showing that under the conditions in Theorem~\ref{thm:m_sufficiently_accurate}, the bound $\bar{b}(\theta, \mu)$ in \eqref{eq:b_A_upper} is indeed the maximum value that $f(\theta^+) - f(\theta) + c\mu\nabla f(\theta)^\intercal \nabla m_k(\theta)$ can obtain given the input $\theta\in\mathbb{R}^{n_\theta}$. That is, we will start by showing that \eqref{eq:b_A_upper} is the same as
\begin{subequations}
    \label{eq:b_A_upper_proof}
    \begin{align}
        \bar{b}(\theta,\mu) =&\nonumber\\
        \underset{f\in\mathcal{H}}{\sup} & \: f(\theta^+) - f(\theta) + c\mu\nabla f(\theta)^\intercal \nabla m_k(\theta) \label{eq:b_A_upper_proof_cost}\\
        \text{s.t.} & \: \lVert f \rVert^2 \leq \Gamma^2, \label{eq:b_A_upper_proof_norm}\\
                    & \: \lvert f(\widetilde{\theta}_i) - \widetilde{y}_i \rvert \leq \bar{\delta} \quad \forall i\in\{1,\dots,N_k\}, \label{eq:b_A_upper_bounds}
    \end{align}
\end{subequations}
where $\mathcal{H}$ denotes the reproducing kernel Hilbert space associated with the kernel $\kappa$. The proof follows steps similar to those taken in the proof of \citet[Theorem~1]{Scharnhorst2023}, i.e., first we show that the search for $f \in \mathcal{H}$ in \eqref{eq:b_A_upper_proof} can be restricted to the finite-dimensional subspace  
\begin{align}
    \mathcal{H}^\parallel := \{f\in\; & \mathcal{H} \mid f\in {\rm span}\{\kappa(\cdot,\widetilde{\theta}_1),\dots, \kappa(\cdot,\widetilde{\theta}_{N_k}), \kappa(\cdot,\theta),\nonumber\\
    &\kappa(\cdot,\theta^+), D^{(0,e_1)}\kappa(\cdot,\theta), \dots, D^{(0, e_{n_\theta})}\kappa(\cdot,\theta)\}\}\label{eq:H_parallel}
\end{align}
with $D^{(0,e_i)}\kappa(\cdot,\theta)$ as in \eqref{eq:kernel_derivatives}. Next, we show that the supremum in \eqref{eq:b_A_upper_proof} can be replaced by a maximum. Finally, we show that from these two observations it follows that \eqref{eq:b_A_upper_proof} can be written as \eqref{eq:b_A_upper}.

\underline{Step~1:} Note that by the reproducing property of the kernel $\kappa$ and its partial derivatives \citep{Zhou2008} we have for any $z\in\mathbb{R}^{n_\theta}$ that
\begin{align}
    f(z) = \langle f, \kappa(\cdot,z) \rangle \label{eq:fz}
\end{align}
and
\begin{align}
    \nabla f(z)
    &= \begin{bmatrix} \langle f, D^{(0, e_1)}\kappa(\cdot, z) \rangle & \dots & \langle f, D^{(0,e_{n_\theta})}\kappa(\cdot,z)\rangle \end{bmatrix}^\intercal.\label{eq:nabla_fz}
\end{align}
Let $\mathcal{H}^\perp := \{g\in\mathcal{H} \mid \langle g, f^\parallel \rangle=0, \, \forall f^\parallel \in \mathcal{H}^\parallel\}$ be the subspace of functions in $\mathcal{H}$ that are orthogonal to $\mathcal{H}^\parallel$ in \eqref{eq:H_parallel}. It then follows that $\mathcal{H} = \mathcal{H}^\parallel \oplus \mathcal{H}^\perp$, with $\oplus$ denoting the vector direct sum, and that for all $f\in\mathcal{H}$ there exist $f^\parallel\in\mathcal{H}^\parallel$ and $f^\perp\in\mathcal{H}^\perp$ such that $f = f^\parallel + f^\perp$. Using this decomposition of $f$, and observing that $\langle f^\perp, \kappa(\cdot, z) \rangle$ and $\langle f^\perp, D^{(0,e_j)}\kappa(\cdot,\theta) \rangle$ are zero for $z\in\{\widetilde{\theta}_1,\widetilde{\theta}_2,\dots,\widetilde{\theta}_{N_k},\theta,\theta^+\}$ and $j=1, 2,\dots,n_\theta$, respectively, since for those $z$ and $j$ it holds that $\kappa(\cdot,z)\in\mathcal{H}^\parallel$, and $D^{{(0,e_j)}}\kappa(\cdot, \theta)\in \mathcal{H}^\parallel$, we obtain using \eqref{eq:fz} and \eqref{eq:nabla_fz} that \eqref{eq:b_A_upper_proof_cost} can be written as
\begin{align}
    f(\theta^+) - &f(\theta) + c\mu\nabla f(\theta)^\intercal \nabla m_k(\theta) \nonumber\\
    &=f^\parallel(\theta^+)-f^\parallel(\theta)+c\mu\nabla f^\parallel(\theta)^\intercal \nabla m_k(\theta)\label{eq:b_A_upper_proof_cost_new}
\end{align}
and that \eqref{eq:b_A_upper_bounds} can be written as
\begin{align}
    \lvert f(\widetilde{\theta}_i) - \widetilde{y}_i \rvert = \lvert f^\parallel(\widetilde{\theta}_i) - \widetilde{y}_i \rvert.\label{eq:b_A_upper_bound_cost_new}
\end{align}
Furthermore, it follows from the same decomposition that
\begin{align}
   \lVert f \rVert^2 =  \langle f, f \rangle =  \langle f^\parallel+f^\perp, f^\parallel+f^\perp \rangle = \lVert f^\parallel \rVert^2 + \lVert f^\perp \rVert^2
    \label{eq:b_A_upper_proof_norm_new}
\end{align}
since $f^\parallel$ and $f^\perp$ are orthogonal. Substituting \eqref{eq:b_A_upper_proof_cost_new}-\eqref{eq:b_A_upper_proof_norm_new} in \eqref{eq:b_A_upper_proof} we obtain 
\begin{subequations}
    \label{eq:b_A_f_parallel}
    \begin{align}
        \bar{b}(\theta, \mu) =&\nonumber\\ 
        \underset{\substack{f^\parallel\in\mathcal{H}^\parallel,\\ f^\perp\in\mathcal{H}^\perp}}{\sup} & \: f^\parallel(\theta^+)-f^\parallel(\theta)+c\mu\nabla f^\parallel(\theta)^\intercal \nabla m_k(\theta) \label{eq:b_A_upper_norm_f_parallel_cost}\\
        \text{s.t.} & \: \lVert f^\parallel \rVert^2 + \lVert f^\perp \rVert^2 \leq \Gamma^2,\label{eq:b_A_upper_norm_f_parallel_norm}\\
        & \: \lvert f^\parallel(\widetilde{\theta}_i) - \widetilde{y}_i \rvert \leq \bar{\delta} \; \forall i\in\{1,2,\dots,N_k\}\label{eq:b_A_upper_norm_f_parallel_bound}.
    \end{align}
\end{subequations}
Note that the objective \eqref{eq:b_A_upper_norm_f_parallel_cost} does not depend on $f^\perp$, and that any $f^\perp\neq 0$ would tighten the constraint \eqref{eq:b_A_upper_norm_f_parallel_norm}. Hence, the supremum (if attained) will be attained for $f^\perp=0$, and thus the search over $f\in\mathcal{H}$ in \eqref{eq:b_A_upper_proof} can be restricted to the finite-dimensional subspace $\mathcal{H}^\parallel$ as in \eqref{eq:H_parallel}.

\underline{Step~2:} Next, we address the attainment of the supremum by showing that the constraints in \eqref{eq:b_A_f_parallel} define a closed and bounded feasible set. First, note that $\lVert f^\parallel \rVert^2\leq \Gamma^2$ defines a closed and bounded set, since it is the sublevel set of a norm. Next, we note that since sets of the form $\{a \in \mathbb{R} \mid \lvert a \rvert \leq b\}$ are closed in $\mathbb{R}$, so are $\{f^\parallel(\widetilde{\theta}_i)\in \mathbb{R} \mid \lvert f^\parallel(\widetilde{\theta}_i) - \widetilde{y}_i \rvert \leq \bar{\delta}\}$ $\forall i\in\{1,2,\dots,N_k\}$. Furthermore, the evaluation functional $L_z(f^\parallel)= f^\parallel(z)$ is a linear operator and thus the pre-images $L_z^{-1}$ of these closed sets are also closed. Consequently, $\{f^\parallel\in \mathcal{H}^\parallel \mid \lvert f^\parallel(\widetilde{\theta}_i) - \widetilde{y}_i \rvert \leq \bar{\delta}\}$ $\forall i\in\{1,2,\dots,N_k\}$ are closed in $\mathcal{H}^\parallel$. Since the intersection of a finite number of closed sets is necessarily closed, the constraints \eqref{eq:b_A_upper_norm_f_parallel_norm}-\eqref{eq:b_A_upper_norm_f_parallel_bound} define a closed feasible set. Moreover, this set is bounded because it is contained in the bounded set defined by the constraint \eqref{eq:b_A_upper_norm_f_parallel_norm}. Since $\mathcal{H}^\parallel$ in \eqref{eq:H_parallel} is finite dimensional, any closed and bounded subset is compact by the Heine-Borel theorem (see, e.g., \citet[Theorem~20]{Royden2010}). Therefore, by the extreme value theorem, the continuous objective \eqref{eq:b_A_upper_norm_f_parallel_cost} attains a maximum on the feasible compact set. Moreover, from Step~1 it follows that the optimizer for which this maximum is obtained must be in $\mathcal{H}^\parallel$ ($f^\perp = 0$), whose members $f^\parallel\in\mathcal{H}^\parallel$ by \eqref{eq:H_parallel} have the form 
\begin{align}
    f^\parallel(\cdot) &= \begin{bmatrix} K_\Omega(\cdot) & \kappa(\cdot,\theta) &\kappa(\cdot,\theta^+) & \nabla K(\theta,\cdot)^\intercal\end{bmatrix}\beta\label{eq:fz_values}
\end{align}
with $K_\Omega(\cdot)$ as in \eqref{eq:m_k_sum} for the inputs $\Omega=\{\widetilde{\theta}_1,\widetilde{\theta}_2,\dots,\widetilde{\theta}_{N_k}\}$, $\nabla K(\theta,\cdot)$ as in \eqref{eq:nabla_K_ab}, and $\beta\in\mathbb{R}^{N_k + 2 + n_\theta}$ a weight vector.

\underline{Step~3:} Finally, we show that it follows from the above that \eqref{eq:b_A_upper_proof} can be written as \eqref{eq:b_A_upper}. To this end, note that it follows from the definition of $\mathfrak{K}^\prime(\theta,\theta^+)$ in \eqref{eq:K_prime_theta}, and \eqref{eq:fz_values} and its partial derivatives that 
\begin{align}
    &\begin{bmatrix} f^\parallel(\widetilde{\theta}_1) &  f^\parallel(\widetilde{\theta}_2) & \cdots & f^\parallel(\widetilde{\theta}_{N_k}) & f^\parallel(\theta) & f^\parallel(\theta^+) &  \nabla f(\theta)^\intercal \end{bmatrix}^\intercal\nonumber\\
    &\hphantom{\begin{bmatrix} f^\parallel(\widetilde{\theta}_1) & \cdots & f^\parallel(\widetilde{\theta}_{N_k}) & f^\parallel(\theta) \end{bmatrix}^\intercal}=\mathfrak{K}^\prime(\theta,\theta^+)\beta\label{eq:f_parallel_D}
\end{align}
Using \eqref{eq:f_parallel_D}, the objective \eqref{eq:b_A_upper_norm_f_parallel_cost} can be written as
\begin{align}
    f^\parallel&(\theta^+)-f^\parallel(\theta) + c\mu\nabla f^\parallel(\theta)^\intercal\nabla m_k(\theta) \nonumber\\
    & =\begin{bmatrix} 0_{1\times N_k}& -1 & 1 & c\mu\nabla m_k(\theta)^\intercal \end{bmatrix}\mathfrak{K}^\prime(\theta,\theta^+)\beta.\label{eq:objective}
\end{align}
Furthermore, using \eqref{eq:fz_values}, the definition of $\mathfrak{K}^\prime(\theta,\theta^+)$ in \eqref{eq:K_prime_theta}, and the reproducing properties \eqref{eq:fz} and \eqref{eq:nabla_fz}, the constraint \eqref{eq:b_A_upper_norm_f_parallel_norm} can be written similar to \eqref{eq:g_norm} as
\begin{align}
    \lVert f^\parallel \rVert^2 
    &= \langle f^\parallel, f^\parallel \rangle
    = \beta^\intercal\mathfrak{K}^\prime(\theta,\theta^+)\beta \leq \Gamma^2,\label{eq:norm_bound_constraint}
\end{align}
while the constraint \eqref{eq:b_A_upper_norm_f_parallel_bound} can be written as
\begin{align}
    \lvert f^\parallel(\widetilde{\theta}_i) - \widetilde{y}_i \rvert = \lvert \xi_i^\intercal\mathfrak{K}^\prime(\theta,\theta^+)\beta - \widetilde{y}_i \rvert \leq \bar{\delta} \label{eq:perturbation_bound_constraint_proof}
\end{align}
for all $i\in\{1,2,\dots,N_k\}$ with $\xi_i\in\mathbb{R}^{N_k+2+n_\theta}$ a vector with $i$-th element equal to one and other elements equal to zero. 

Using \eqref{eq:objective}-\eqref{eq:perturbation_bound_constraint_proof}, and the observation from Step 2 that the objective \eqref{eq:b_A_upper_norm_f_parallel_cost} attains a maximum on the feasible compact set, we obtain \eqref{eq:b_A_upper} from \eqref{eq:b_A_f_parallel}, completing the proof for $\bar{b}(\theta,\mu)$. 

The proof that under the conditions in Theorem~\ref{thm:m_sufficiently_accurate} $\underaccent{\bar}{b}(\theta)$ in \eqref{eq:b_nabla_upper} is a lower bound on $\nabla f(\theta)^\intercal \nabla m_k(\theta)$, i.e., that \eqref{eq:b_nabla_upper} is the same as
\begin{subequations}
    \label{eq:b_nabla_upper_proof}
    \begin{align}
        \underaccent{\bar}{b}(\theta) =&\nonumber\\
        \underset{f\in\mathcal{H}}{\inf} \: &\nabla f(\theta)^\intercal \nabla m_k(\theta) \\
        \text{s.t.} \: & \lVert f \rVert^2 \leq \Gamma^2,\\
        & \lvert f(\widetilde{\theta}_i) - \widetilde{y}_i \rvert \leq \bar{\delta}\quad \forall i\in\{1,2,\dots,N_k\},
    \end{align}
\end{subequations}
follows \emph{mutatis mutandis} from the above proof by omitting $\kappa(\cdot,\theta)$ and $\kappa(\cdot,\theta^+)$ from the span in \eqref{eq:H_parallel}, using $\mathfrak{K}(\theta)$ as in \eqref{eq:K_theta} instead of $\mathfrak{K}^\prime(\theta,\theta^+)$ as in \eqref{eq:K_prime_theta}, removing $f(\theta^+)$, $f(\theta)$, $c$, and $\mu$ from the objective \eqref{eq:b_A_upper_proof_cost}, and taking the infimum instead of the supremum.
\section{Proof of Lemma~\ref{lem:upper_bound_DV_kernel-based}}
\label{app:proof_upper_bound_DV_kernel-based}

The proof of Lemma~\ref{lem:upper_bound_DV_kernel-based} is as follows. Since by the conditions of the lemma $\bar{b}(\theta,\mu) \leq 0$, it follows from Theorem~\ref{thm:m_sufficiently_accurate} that the inequality $f(\theta^+) \leq f(\theta) - c \mu \nabla f(\theta)^\intercal \nabla m_k(\theta)$ with $c \in (0, 1)$ is satisfied. Therefore, using the definitions of $V(\cdot)$ and $\Delta V(\cdot)$ in \eqref{eq:definition_V} and \eqref{eq:definition_DV}, we obtain that the inequality
\begin{align}
    \Delta V(\theta) &= f(\theta^+) - f(\theta) \leq - c \mu \nabla f(\theta)^\intercal \nabla m_k(\theta) \label{eq:DV_nabla_f_nabla_m}
\end{align}
holds. Furthermore, since by the conditions of the lemma $\underaccent{\bar}{b}(\theta) > 0$ for all $\theta \in \mathbb{R} \backslash \mathcal{C}$, and by Assumption~\ref{as:minimum_f} $\nabla f(\theta) = 0$ if and only if $\theta \in \mathcal{C}$, it follows from Theorem~\ref{thm:m_sufficiently_accurate} and \eqref{eq:DV_nabla_f_nabla_m} that $\Delta V(\theta) < 0$ for all $\theta \in \mathbb{R}^{n_\theta} \backslash \mathcal{C}$ and $\Delta V(\theta) = 0$ for all $\theta \in \mathcal{C}$. Hence, for any $\Delta_\theta$ such that $\lVert \theta \rVert_{\mathcal{C}} \leq \Delta_\theta$, there exists a class-$\mathcal{K}$ function $\bar{\rho}_k$ such that $\bar{\rho}_k(\lVert \theta \rVert_{\mathcal{C}}) \leq -\Delta V(\theta)$ (see, e.g., \citet[Lemma~4.3]{Khalil2002}). Moreover, it follows from Assumption~\ref{as:DV_standard} that $\lVert \theta \rVert_{\mathcal{C}} \geq \omega_2^{-1}(V(\theta))$ and thus that $-\bar{\rho}_k(\lVert \theta \rVert_{\mathcal{C}}) \leq -\bar{\rho}_k(\omega_2^{-1}(V(\theta)))$. Therefore, we obtain from \eqref{eq:DV_nabla_f_nabla_m} that during kernel-based extremum-seeking steps the inequality 
\begin{align}
    \Delta V(\theta) \leq - \bar{\rho}_k(\lVert \theta \rVert_{\mathcal{C}}) \leq  -\widetilde{\rho}_k(V(\theta)), \label{eq:DV_kb}
\end{align}
with $\widetilde{\rho}_k := \bar{\rho}_k(\omega_2^{-1}(\cdot))$ a function of class $\mathcal{K}$, holds for all $\theta$ such that $\lVert \theta \rVert_{\mathcal{C}}\leq \Delta_\theta$, which completes the proof.
\section{Proof of Theorem~\ref{thm:stability_kernel-based_ESC}}
\label{app:proof_stability_kernel-based_ESC}

The proof of Theorem~\ref{thm:stability_kernel-based_ESC} consists of three steps. In the first two steps of the proof the increments of Lemmas~\ref{lem:upper_bound_DW_standard} and \ref{lem:upper_bound_DV_kernel-based} are used to derive upper bounds on $W(\psi, x, \theta)$ after a sequence of, respectively, standard or kernel-based update steps. In the third step of the proof, these upper bounds are combined to determine an upper bound that holds for either type of update step, from which it can be concluded that $\lVert \theta \rVert_{\mathcal{C}}$ ultimately converges to the level set of $\lVert \cdot \rVert_\mathcal{C}$ given in the theorem.

\underline{Step 1:} We first show that from Lemma~\ref{lem:upper_bound_DW_standard} it follows that $W(\psi, x, \theta)$ is upper bounded during standard update steps by the maximum of a class-$\mathcal{K}\mathcal{L}$ function and a constant. To this end, let $\widehat{\rho}$ be a class-$\mathcal{K}_\infty$ function such that $({\rm id}-\widehat{\rho}) \in \mathcal{K}_\infty$. By Lemma~\ref{lem:upper_bound_DW_standard}, it then follows that for any $\theta, \psi \in \mathbb{R}^{n_\theta}$ and $x \in \mathcal{X}$ such that $\lVert \theta \rVert_{\mathcal{C}}\leq \Delta_\theta$, $\lVert \psi \rVert_{\mathcal{C}} \leq \Delta_\theta$, and $\lVert x \rVert_{\mathcal{A}(\psi + v_{n_v}(\theta))} \leq \Delta_x$, the inequality 
\begin{align}
    \Delta W(\psi, x, \theta) &\leq -\widetilde{\gamma} \left( W(\psi, x, \theta) \right) + \gamma + 2\delta + \delta_V\\
    &= -({\rm id} - \widehat{\rho}) \circ \widetilde{\gamma} \left( W(\psi, x, \theta)\right) + \gamma + 2\delta \nonumber \\
    &\hphantom{=~} + \delta_V -\widehat{\rho} \circ \widetilde{\gamma} \left(W(\psi, x, \theta)\right)\\
    &\leq -({\rm id} - \widehat{\rho}) \circ \widetilde{\gamma} \left( W(\psi, x, \theta) \right)\label{eq:DW_id_rhohat_standard}
\end{align}
holds as long as $\widehat{\rho} \circ \widetilde{\gamma}(W(\psi, x, \theta)) \geq \gamma + 2\delta + \delta_V$. From  \eqref{eq:DW_id_rhohat_standard} and the comparison lemma (see, e.g., \citet[Lemma~4.3]{Jiang2002}), it follows that there exists a class-$\mathcal{K}\mathcal{L}$ function $\beta_1$ such that
\begin{align}
    W(\psi_{k + \ell}, x_{k + \ell}, \theta_{k + \ell}) \leq \max\{&\beta_1(W(\psi_k, x_k, \theta_k), \ell),\nonumber \\
    & \; \varphi(\gamma + 2\delta + \delta_V)\} \label{eq:W_beta1}
\end{align} 
for a sequence of $\ell$ standard update steps starting at any update number $k$, with $\varphi(\cdot) := \widetilde{\gamma}^{-1} \circ \widehat{\rho}^{-1}(\cdot)$ a function of class $\mathcal{K}_{\infty}$. 

\underline{Step 2:} Next, note that a kernel-based update \eqref{eq:theta_plus_kb} is only performed if $\underaccent{\bar}{b}(\theta) > 0$ and with $\mu \in \mathbb{R}_{> 0}$ chosen such that $\bar{b}(\theta, \mu) \leq 0$ (cf. Algorithm~\ref{alg:kb_esc}). Therefore, it follows from Lemma~\ref{lem:upper_bound_DV_kernel-based} that 
\begin{align}
    \Delta V(\theta) \leq -\widetilde{\rho}_k(V(\theta))
\end{align}
for any $\theta \in \mathbb{R}^{n_\theta}$ such that $\lVert \theta \rVert_{\mathcal{C}}  \leq \Delta_\theta$ at any kernel-based update step $k$. Therefore, for a sequence of $\ell$ kernel-based update steps, starting at update step $k$, it holds that 
\begin{align}
    \Delta V(\theta) \leq -\underaccent{\bar}{\rho}_{k,\ell}(V(\theta)) \label{eq:DV_rho_tilde}
\end{align}
at every step, where $\underaccent{\bar}{\rho}_{k,\ell}(\cdot) := \min\{\widetilde{\rho}_k(\cdot), \dots, \widetilde{\rho}_{k+\ell}(\cdot)\}$ is a function of class $\mathcal{K}$. Similar to Step 1, we conclude from \eqref{eq:DV_rho_tilde} and the comparison lemma that there exists a class-$\mathcal{K}\mathcal{L}$ function $\beta_2$ such that 
\begin{align}
    V(\theta_{k + \ell}) \leq \beta_2(V(\theta_k), \ell) \label{eq:V_beta2}
\end{align}
for a sequence of $\ell$ kernel-based update steps starting at any update number $k$. Substituting \eqref{eq:V_beta2} in the definition of $W(\psi, x, \theta)$ in \eqref{eq:definition_W}, we obtain that
\begin{align}
    W&(\psi_{k+\ell}, x_{k+\ell}, \theta_{k + \ell}) \leq V_p(\psi_{k + \ell}, x_{k + \ell}) + 2\beta_2(V(\theta_k), \ell)\label{eq:W_beta2}
\end{align}
during a sequence of $\ell$ kernel-based update steps starting from update number $k$. Note that during this sequence, the memory state $\psi$ and the system state $x$ do not change, since the input applied to the system $\Sigma_p$ is not changed during kernel-based update steps (cf. Algorithm~\ref{alg:kb_esc}, Line~\ref{line:kb_update}), and since kernel-based update steps are instantaneous (because there is no waiting time $T$ for these steps).

\underline{Step 3:} Finally, we combine the observations from Steps 1 and 2 to draw conclusions about the level set of $\lVert \cdot \rVert_\mathcal{C}$ to which $\lVert \theta \rVert_{\mathcal{C}}$ converges. To this end, note that it follows from \eqref{eq:W_beta1} and \eqref{eq:W_beta2} that 
\begin{align}
    W(& \psi_{k + \ell}, x_{k + \ell}, \theta_{k + \ell}) \nonumber \\
    & \leq \max\{\beta_1(W(\psi_k, x_k, \theta_k), \ell), \, \varphi(\gamma + 2\delta + \delta_V), \nonumber \\
    & \hphantom{\leq~\max\{}\; V_p(\psi_{k + \ell}, x_{k + \ell}) + 2\beta_2(V(\theta_k), \ell)\} \label{eq:W_beta_combined}
\end{align}
for a sequence of $\ell$ update steps of either type starting from update number $k$. Subtracting $V_p(\psi_{k + \ell}, x_{k + \ell})$ from both sides in \eqref{eq:W_beta_combined} and noting that $V_p(\psi,x) \geq 0$, it thus follows using the definition of $W(\psi, x, \theta)$ in \eqref{eq:definition_W} that
\begin{align}
    2V(\theta_{k + \ell}) \leq \max\{&\beta_1(W(\psi_k, x_k, \theta_k), \ell), \, 2\beta_2(V(\theta_k), \ell), \nonumber \\
    &\; \varphi(\gamma + 2\delta + \delta_V)\}. \label{eq:V_beta_combined}
\end{align}
Since both $\beta_1$ and $\beta_2$ are functions of class $\mathcal{K}\mathcal{L}$, it follows from \eqref{eq:V_beta_combined} that starting from the initial update $k=0$, $V(\theta)$ is ultimately bounded by
\begin{align}
    \lim_{\ell \to \infty} V(\theta_\ell) \leq \frac{1}{2}\varphi(\gamma + 2\delta + \delta_V).\label{eq:ultimate_bound_V}
\end{align}
Finally, since $\omega_1(\lVert \theta \rVert_{\mathcal{C}}) \leq V(\theta)$ by Assumption~\ref{as:DV_standard}, we obtain from \eqref{eq:ultimate_bound_V} that the optimizer state $\theta$ ultimately converges to the set
\begin{align}
    \left\{\theta\in\mathbb{R}^{n_\theta} \mid \lVert \theta \rVert_{\mathcal{C}} \leq \omega_1^{-1}\left(\frac{1}{2}\varphi(\gamma+2\delta+\delta_V)\right)\right\}.
\end{align}
Since $\gamma$, $\delta$, and $\delta_V$ can be made arbitrarily small by Lemma~\ref{lem:upper_bound_DW_standard}, this set to which $\theta$ converges can be made arbitrarily small, completing the proof.

\bibliographystyle{elsarticle-harv}
\bibliography{references}
\end{document}